\documentclass[preprintnumbers, floatfix,letterpaper, twocolumn,aps,prd,epsfig,nofootinbib]{revtex4-1}
\usepackage{bm, graphicx, dcolumn, epstopdf, epsf, latexsym, mathbbol, amssymb, amsmath, slashed, mathrsfs, mathcomp, simplewick, color, appendix}
\usepackage[center]{subfigure}
\usepackage{multirow}
\usepackage{makecell}
\usepackage[colorlinks,linkcolor=blue,citecolor=blue,urlcolor=blue]{hyperref}

\begin{document}
\allowdisplaybreaks
 \newcommand{\bq}{\begin{equation}}
 \newcommand{\eq}{\end{equation}}
 \newcommand{\bqn}{\begin{eqnarray}}
 \newcommand{\eqn}{\end{eqnarray}}
 \newcommand{\nb}{\nonumber}
 \newcommand{\lb}{\label}
 \newcommand{\f}{\frac}
 \newcommand{\p}{\partial}
\newcommand{\PRL}{Phys. Rev. Lett.}
\newcommand{\PLB}{Phys. Lett. B}
\newcommand{\PRD}{Phys. Rev. D}
\newcommand{\CQG}{Class. Quantum Grav.}
\newcommand{\JCAP}{J. Cosmol. Astropart. Phys.}
\newcommand{\JHEP}{J. High. Energy. Phys.}
\title{High-order Primordial Perturbations with Quantum Gravitational Effects}

\author{Tao Zhu${}^{a, b}$}
\email{Tao$\_$Zhu@baylor.edu} 

\author{Anzhong Wang${}^{a, b, c}$}
\email{anzhong$\_$wang@baylor.edu} 

\author{Klaus Kirsten${}^{d}$}
\email{klaus$\_$kirsten@baylor.edu} 

\author{Gerald Cleaver${}^{e}$}
\email{gerald$\_$cleaver@baylor.edu} 

\author{Qin Sheng${}^{d}$}
\email{qin$\_$sheng@baylor.edu} 

\affiliation{${}^{a}$ Institute for Advanced Physics $\&$ Mathematics, Zhejiang University of Technology, Hangzhou, 310032, China\\
${}^{b}$ GCAP-CASPER, Physics Department, Baylor University, Waco, TX 76798-7316, USA\\
$^{c}$ Departamento de F\'{\i}sica Te\'orica, Instituto de F\'{\i}sica, UERJ, 20550-900, Rio de Janeiro, Brazil\\
${}^{d}$ GCAP-CASPER, Mathematics Department, Baylor University, Waco, TX 76798-7328, USA\\
${}^{e}$ EUCOS-CASPER, Physics Department, Baylor University, Waco, TX 76798-7316, USA}

\date{\today}

\begin{abstract}
In this paper, we provide a systematic investigation of high-order primordial perturbations with nonlinear dispersion relations due to quantum gravitational effects in the framework of {\em uniform asymptotic approximations}. Because of these effects, the equation of motion of the mode function in general has multiple-turning points. After obtaining analytically approximated solutions to any order in different regions, associated with different types of turning points, we match them to the third one. To this order the  errors are less than $0.15\%$.  General expressions of the power spectra of the primordial tensor and scalar perturbations are derived explicitly. We also investigate effects of back-reactions of the quantum gravitational corrections, and make sure that inflation lasts long enough in order to solve the underlying problems, such as flatness, horizon and monopole. Then, we study various features of the spectra that are observationally relevant. In particular, under a moderate assumption about the energy scale of the underlying theory of quantum gravity, we have shown that the quantum gravitational effects may alter significantly the ratio between the tensor and scalar power spectra, thereby providing a natural mechanism to alleviate the tension between observations and certain inflationary models, including the one with a quadratic potential.
\end{abstract}
\maketitle

\section{Introduction}
\renewcommand{\theequation}{1.\arabic{equation}}
\setcounter{equation}{0}

The paradigm of cosmic inflation has had remarkable success in solving many problems elegantly 
with standard big bang cosmology and predicting the primordial power spectrum whose evolution 
explains both the formation of the large scale structure of the universe and the small inhomogeneities 
in the cosmic microwave background (CMB) \cite{Guth}. These are now matched to observations with 
a spectacular precision \cite{WMAP, PLANCK2013, PLANCK2015}. However, as it is well known, the 
inflationary scenario is conceptually incomplete in several respects. For example, in some inflation models, 
the energy scale of quantum fluctuations, which relates to the present observations, was not far from the 
Planck energy in the beginning of the inflation \cite{Brandenberger1999, Martin2001}.  This leads to the 
problem that the underlying quantum field theory on classical spacetime becomes unreliable. In addition, 
the evolution of the background geometry and matter, which satisfies Einstein's equation during the 
inflational process, will inevitably lead to the existence of an initial singularity 
\cite{Brandenberger2013CQG,bigbangsingularity}. Yet, the inflation paradigm in general ignores the 
pre-inflationary dynamics and simply sets the initial condition of the perturbations in the Bunch-Davies 
vacuum at the onset of the slow-roll epoch. To resolve any of these problems, the ultraviolet (UV) physics is
needed. 

So far, various approaches have been proposed to study the aforementioned effects; see, for example,  
\cite{Brandenberger1999,bigbangsingularity,Martin2001, Brandenberger2013CQG, eff, UCJ,JM09}, and 
references therein. In most of such considerations, both scalar and tensor perturbations produced during 
the inflationary epoch are governed by the equation
\bqn\lb{eom}
\mu_k''(\eta)+\left(\omega_k^2(\eta)-\frac{z''}{z}\right)\mu_k(\eta)=0,
\eqn
where $\mu_k(\eta)$ denotes the mode function of inflationary perturbations, and a prime indicates 
differentiation with respect to the conformal time $\eta.$ $k$ is the comoving wave number, and $z=z(\eta)$ 
depends on the background and the types of perturbations (scalar and tensor). In the present paper, we 
will consider the particular nonlinear dispersion relation
\bqn\lb{nonlinear}
\omega^2_k(\eta)=k^2 \left[1-\hat b_1\left(\frac{k}{aM_*}\right)^2+\hat b_2 \left(\frac{k}{aM_*}\right)^4 \right],
\eqn
where $M_*$ is the relevant energy scale of the nonlinear dispersion relation, $a=a(\eta)$ is the scale factor 
of the background universe, and $\hat b_1$ and $\hat b_2$ are dimensionless constants. In order to get a 
healthy ultraviolet limit if $\hat b_1 \neq 0,$ one in general requires $\hat b_2>0.$ When $\hat b_1=0=\hat b_2$, 
it reduces to that of general relativity. 

The nonlinear dispersion relation in Eq.~(\ref{nonlinear}) was first suggested for inflationary perturbations in \cite{Brandenberger1999} as a toy model, which has been used to study the unknown effects of trans-Planckian physics. Later it was found that it could be naturally realized in the framework of Ho\v{r}ava-Lifshitz theory of gravity \cite{HL, InfHL}. While the last two terms on the right hand side of Eq.~(\ref{nonlinear}) denote the contributions from the fourth and sixth-order spatial operators in the theory of Ho\v{r}ava-Lifshitz gravity, the term $\omega_k^2$ comes from the fact that the theory only allows the second-order time derivative operators. The essential point of keeping the time derivative operators to the second-order and the high-dimensional spatial operators up to the sixth-order is that it not only reserves the unitarity of the theory, but also makes the theory power-counting renormalizable. 

To have a better  physical understanding of  the quantum gravitational effects, it is highly desirable  to solve Eq.(\ref{eom}) analytically for the nonlinear dispersion relation   (\ref{nonlinear}), and then extract information from it, including the power  spectra, spectral indices and runnings. Such studies are very challenging, as the problem becomes so mathematically involved, and meanwhile the treatment needs to be sufficiently accurate, in order to identify precisely such effects and seek  observational signatures in the current and forthcoming experiments. Recently, we  developed a powerful and effective method, {\em the uniform asymptotic approximation} to accurately construct analytical solutions of the mode function with a nonlinear dispersion relation \cite{uniform_nonlinear}. It should be noted that the uniform asymptotic approximation was first applied to inflationary cosmology by Habib {\em et al.} \cite{uniformPRL}, in which the dispersion relation is the standard linear relation $\omega_k(\eta)=k^2$, where the equation of motion only has one single turning point  (see also \cite{HL}).  We generalized the method of Habib {\em et al.} to the case with the nonlinear dispersion relation of Eq.(\ref{nonlinear}), where multiple and high-order turning points are allowed. Furthermore, we extended the first-order uniform asymptotic approximation to the third-order  in cases where there exists only one-turning point \cite{uniform_singleH}. The formulas  were then applied to $k$-inflationary models  \cite{uniformK}  as well as   to
the ones  in loop quantum cosmology (LQC) with holonomy and inverse-volume  corrections \cite{uniform_loop}.

The aim of this  paper is to generalize our  studies carried out in \cite{uniform_nonlinear} from 
the first-order to the third-order approximations with any number and order of turning-points for the dispersion 
relation given by Eq.(\ref{nonlinear}). Up to the third-order, the upper bounds of errors are $\lesssim 0.15\%$ 
\cite{uniform_singleH}, which are accurate enough for the current and forthcoming experiments \cite{S4-CMB}.
We emphasize that the method to be developed in this paper is quite general, and can easily be 
extended to other inflationary models with different quantum gravitational effects. To proceed further,  we assume that the 
background evolution with an inflationary scalar field satisfies the usual Friedmann and Klein-Gordon equations
\footnote{This assumption is not essential, and can be easily generalized to the case where the 
evolution of the background is affected by quantum gravitational corrections. Such corrections could lead to 
non-slow-roll inflation, such as the case in LQC \cite{uniform_loop}. However,  the uniform asymptotic approximation
method is not restricted to slow-roll inflationary models, and can be thus in principle equally applied to the non-slow-roll case.}, 
\bqn
H^2=\frac{1}{3M_{\text{Pl}}^2} \left(\frac{1}{2}\dot \phi^2+V(\phi)\right),\\
\ddot \phi+3 H \dot \phi+\frac{dV(\phi)}{d\phi}=0,
\eqn
where  $H\equiv \dot a /a$ is the Hubble parameter, $V(\phi)$ is the potential of the inflation field $\phi$, and $M_{\text{Pl}}$ is 
the reduced Planck mass. In the above, a dot represents the derivative with respect to the cosmic time $t$, 
which in terms of the conformal time is given as
\bqn
\eta(t) =\int_{t_{\text{end}}}^t \frac{dt'}{a(t')}.
\eqn
Here $t_{\text{end}}$ is the cosmic time when the inflation ends.

In addition, an important feature of the nonlinear dispersion relation is that it could produce additional excited states for primordial 
perturbations on the sub-horizon scale during inflation.  This has been extensively studied in \cite{backreaction, back-reaction}. 
In general, the solution of Eq.(\ref{eom}) has the following form in the sub-horizon regime $H\ll k/a \ll M_*$ (in the leading WKB approximation),
\bqn\lb{wkb_sol}
\mu_k(\eta) =\frac{1}{\sqrt{2 \omega_k(\eta)}} \left( \tilde\alpha_k e^{-i \int \omega_k d\eta}+\tilde \beta_k e^{i \int \omega_k d\eta}\right),  ~~~~~~~~
\eqn
where $\tilde \alpha_k$ and $\tilde \beta_k$ satisfy the relation
\bqn
|\tilde \alpha_k|^2-|\tilde \beta_k|^2=1.
\eqn
Usually, $\tilde \beta_k=0$ if the mode starts at the Bunch-Davies vacuum and the coefficient term $\Omega_k^2(\eta)=\omega_k^2(\eta)-z''/z$ in Eq.~(\ref{eom}) satisfies the adiabatic condition
\bqn\lb{ad_condition}
\left|\frac{3 (\Omega_k')^2}{4\Omega_k^4} - \frac{\Omega_k''}{2 \Omega_k^3} \right| \ll 1,
\eqn
from the onset of inflation until the sub-horizon regime. However, if the adiabatic condition is violated in the region when  $k /a \gtrsim M_*$, then the primordial perturbation modes through this region will lead to a nonzero $\tilde \beta_k$.  This immediately raises the question whether the back-reaction of such additional excited state (i.e., $\tilde \beta_k$ mode) is small enough to allow inflation to last long enough. Such a consideration leads to the following constraint \cite{back-reaction},
\bqn\lb{constriantAa}
|\tilde \beta_k|^2 \ll 8\pi \frac{M_{\text{Pl}}^2 H_{\text{inf}}^2 }{M_*^4},
\eqn
where $H_{\text{inf}}$ is the Hubble constant when the inflation just starts. Note that similar constraints can also be found 
in \cite{back-reaction2}. In the current  paper, with the  approximate solutions  obtained by using the high-order uniform asymptotic approximation 
to be developed below, we are able to determine the coefficient $\tilde \beta_k$ explicitly, so that we can estimate the effects of the back-reactions 
on the primordial spectra precisely. 

The rest of this paper is organized as follows. In Appendix A, we systematically develop the high-order uniform asymptotic approximation method for multi-turning points, and expand the mode function in terms of $1/\lambda$ to an arbitrarily high order, in which the error bounds are given in each order of the approximation, where $\lambda \gg 1$. Then in Appendix B, we consider the matching of individual approximate solutions constructed in Appendix A. By applying the formalism developed in Appendices A and B to inflationary perturbation modes with the above nonlinear dispersion relation, in Sec. II, we obtain the general expressions of the power spectra to the third-order from the analytically approximate mode function obtained in the uniform asymptotic approximation. Then, we investigate the constraints on the Bogoliubov coefficient $\tilde \beta_k$,  in order for the inflation to last longer enough.   In Sec. III, we investigate features of the power spectra, and study their potential applications to certain inflationary models. In particular, we show that, even with the aforementioned constraints, the model with a quadratic potential can be reconciled with observations after the quantum gravitational corrections are taken into account. Our main conclusions are summarized in Sec. IV.

\section{spectra of primordial perturbations}
\renewcommand{\theequation}{2.\arabic{equation}}
\setcounter{equation}{0}

\subsection{Turning Points  and WKB Approximation}

\begin{figure}
\includegraphics[width=8.1cm]{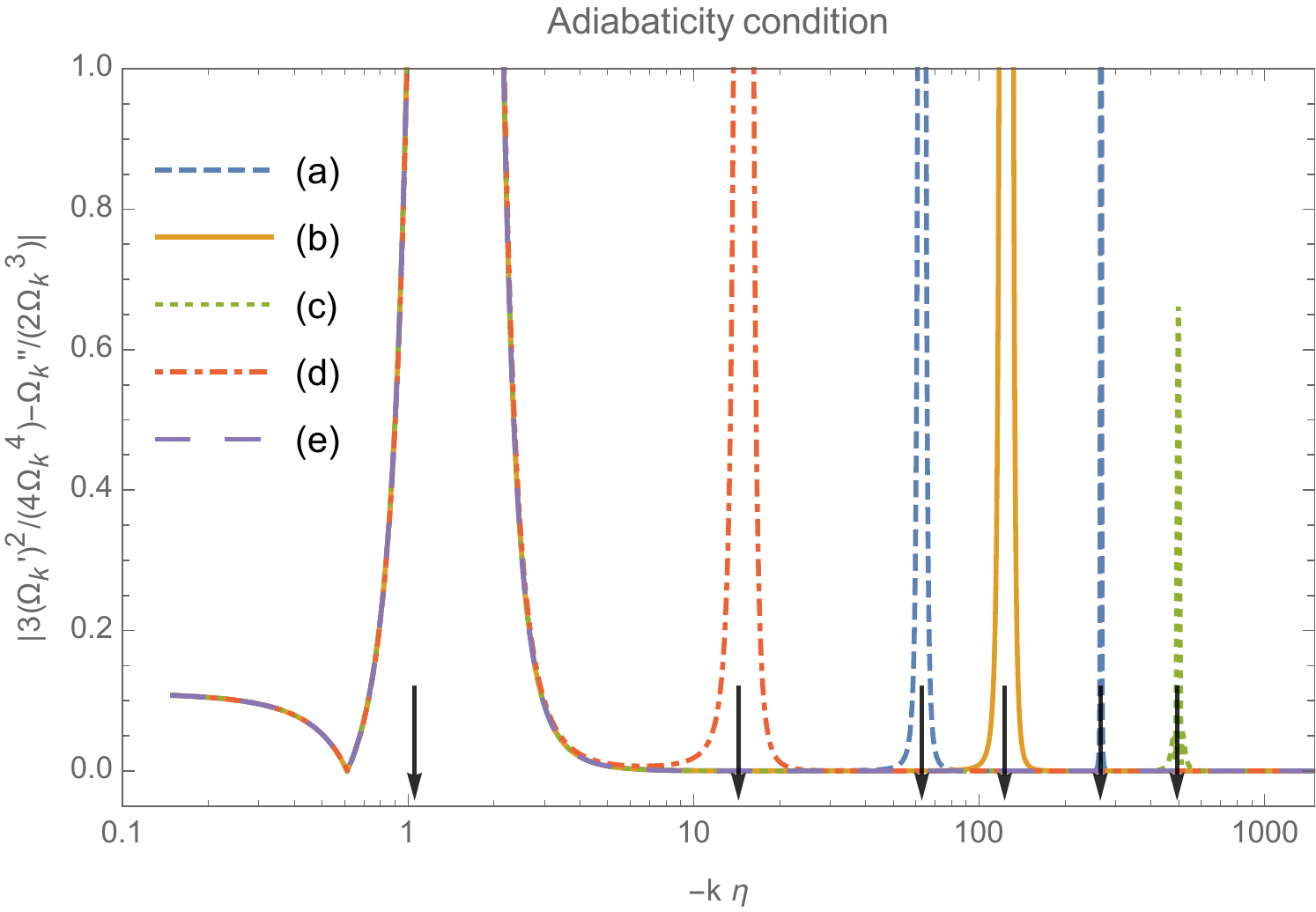}
\caption{The adiabatic condition  of Eq.~(\ref{ad_condition}) for the WKB approximation is violated at the turning points $\Omega_k^2(\eta) \equiv \omega_k^2(\eta)-z''/z = 0$, denoted by the down-pointing vertical arrows. The curves  (a), (b), (c), (d) and (e) correspond, respectively, to the cases illustrated in Fig. \ref{fig_gofy}.}\label{WKB}
\end{figure}

The evolutions of both scalar and tensor perturbations, which obey the equation of motion (\ref{eom}), depend on both the background of the Universe and the nonlinear dispersion relations (\ref{nonlinear}). With the assumption of the slow-roll evolution of the background, then the behaviors of the  mode are very sensitive to the signs of  $\omega^2_k(\eta)-z''/z$ in Eq.~(\ref{eom}). When $\omega_k^2(\eta)\gg |z''/z|$, the solution  is of the type of oscillation, 
\bqn\lb{}
\mu_k(\eta) \sim e^{\pm i \int \omega_k(\eta)d\eta},
\eqn
and when $\omega_k^2(\eta)\ll |z''/z|$, it is of the form, 
\bqn
\mu_k(\eta) \sim \frac{\alpha_1}{z(\eta)}+ \alpha_2 z(\eta).
\eqn
Here we note that the former  corresponds to solutions in the region where the adiabatic condition (\ref{ad_condition}) is fulfilled. Then, the problem of obtaining the solution of Eq.~(\ref{eom}) thus reduces to solving the equation in the intermediate regions when $\omega_k^2(\eta) \sim |z''/z|$. In these regions, the signs of  $\Omega_k^2(\eta) \left[\equiv \omega_k^2(\eta)-z''/z\right]$  change. As a result,  the solution will change from   that of oscillation  to that of   decaying/growing, or vice versa. The transition point where $\Omega_k^2(\eta)= 0$ is usually   called a {\em turning point}. As shown above, the solution of the mode function has  quite different behaviors in different sides of the turning point.  
In particular, in the region where $\Omega_k^2(\eta) \simeq 0$, the adiabatic condition (\ref{ad_condition}) is broken down,  as shown explicitly in Fig. \ref{WKB}, and then the usual WKB approximation becomes invalid.  Therefore, in order to determine the solution of Eq.~(\ref{eom}) in the whole phase space, other approximations near the turning points  are required. To this purpose, in the following we shall show how one can solve   Eq.~(\ref{eom}) at these turning points by using the uniform asymptotic approximation. 

\subsection{Approximate Solutions near Turning Points}

The uniform asymptotic approximation provides a systematic and effective method to construct accurate solutions  near  turning points. To show this, let us re-write Eq.(\ref{eom}) in the form  \cite{Olver1974, Olver1975},
\bqn\lb{standard}
\frac{d^2\mu_k(y)}{dy^2}=\Big[\lambda^2 \hat{g}(y)+q(y)\Big]\mu_k(y),
\eqn
where the new variable $y (\equiv - k\eta)$ is dimensionless, and 
\bqn
\lb{gq}
\lambda^2 \hat g(y)+q(y)&\equiv& -\frac{1}{k^2} \left(\omega_k^2(\eta)-\frac{z''}{z}\right) \nb\\
&=&\frac{\nu^2-1/4}{y^2}-1+b_1 \epsilon_*^2 y^2 -b_2 \epsilon_*^4 y^4,\nb\\
\eqn
where  $\nu^2 \equiv 1/4+\eta^2 z''/z$, $b_1=\hat b_1 /(a^2 H^2 \eta^2)$, $b_2=\hat b_2 /(a^4H^4 \eta^4)$, and $\epsilon_*=H/M_*$. In the above, the quantity $z=z(\eta)$ depends on the types of perturbations.
In particular,  for scalar perturbations we have $z_s(\eta)=a \dot \phi /H$ and for tensor perturbations we have $z_t(\eta)=a$. Restricting to the slow-roll evolution of the background, as we have assumed in this paper, we have $b_1=(1+2 \epsilon_1+\mathcal{O}(\epsilon^2))\hat b_1$, $b_2 =(1+4 \epsilon_1+\mathcal{O}(\epsilon^2))\hat b_2$, $\nu=3/2+\epsilon_1+\epsilon_2/2+\mathcal{O}(\epsilon^2)$ for scalar perturbations, and $\nu=3/2+\epsilon_1+\mathcal{O}(\epsilon^2)$ for tensor perturbations with $\epsilon_1\equiv -\dot H/H^2$, $\epsilon_2\equiv d\ln \epsilon_1/d \ln a$, where $\epsilon_i$ denote the slow-roll  parameters at the first-order slow-roll approximation.

Note that Eq.~(\ref{gq}) cannot determine the two functions $g(y)$ and $q(y)$ uniquely. A fundamental reason to introduce two of them is to have one extra degree of freedom, so later we are allowed to choose them in such a way that the error control functions, associated with the uniform asymptotic approximation, can be minimized. The convergence of these error control functions are very sensitive to their behaviors near the  poles (singularities) of the function $\lambda^2 \hat g(y)$, as we have discussed in details in Appendix \ref{appendixA}. From the expression of Eq.~(\ref{gq}), one can see that there are    two poles, one is located at $y=0^+$ and the other is at $y=+\infty$. Now one can follow the guidance summarized in the last paragraph in Appenxia.~\ref{appendixA} to determine the functions $\lambda^2 \hat g(y)$ and $q(y)$. 

The guidance to determine the functions $\lambda^2 \hat g(y)$ and $q(y)$  contains three conditions. The first condition requires that near  the two poles we must have $|q(y)|<|\lambda^2 \hat g(y)|$, which in turn requires that  $\lambda^2 \hat g(y)$ must be of  the second-order at the pole $y=0^+$, and of  the fourth-order at the pole $y=+\infty$. Then, the second condition requires  that the functions $\lambda^2 \hat g(y)$ and $q(y)$ must be chosen as
\bqn
q(y)&=&- \frac{1}{4 y^2}+\frac{q_1}{y}+q_2+q_3 y+q_4 y^2 +q_5 y^3+q_6 y^4,\nb\\
\lambda^2 \hat g(y)&=&\frac{\nu^2}{y^2}-\frac{q_1}{y}-(1+q_2)-q_3 y\nb\\
&&+(b_1 \epsilon_*^2-q_4) y^2-q_5 y^3 -(b_2 \epsilon_*^4 +q_6)y^4,
\eqn
where $q_i$ with $i=1, 2, 3, 4, 5, 6$ are constants. Considering the third condition one concludes that the pole $y=+\infty$ of $q(y)$ must have order less than one, thus $q_i=0$ with $i>2$. One can choose $q_{1,2}=0$ for the sake of simplicity. Then, 
  the functions $\lambda^2 \hat g(y)$ and $q(y)$ finally take the form
\bqn
q(y)&=&-\frac{1}{4y^2},\\
\lambda^2 \hat g(y) &=&\frac{\nu^2}{y^2}-1+b_1 \epsilon_*^2 y^2-b_2 \epsilon_*^4 y^4. \lb{ggofy}
\eqn
Clearly,  the function $\lambda^2 \hat g(y)$ is free of singularities within the range  $ y\in (0,+\infty)$. 

Except the two poles at $y=0^+$ and $y=+\infty$, $\lambda^2 \hat g(y)$ in general has three zeros for $ y \in(0,+\infty)$, which depends  on the three parameters $b_1$, $b_2$, and $\epsilon_*$. The zeros of $\lambda^2 \hat g(y)$ are also called turning points of the second-order differential equation (\ref{standard}). These turning points can be determined by solving the equation $\lambda^2 \hat g(y)=0$, which are,
\bqn
y_0&=&\left\{\frac{b_1}{3b_2 \epsilon_*^2 }\left[1-2\sqrt{1-\mathcal{Y}}\cos{\left(\frac{\theta}{3}\right)}\right]\right\}^{1/2},\nb\\
y_1&=&\left\{\frac{b_1}{3b_2 \epsilon_*^2 }\left[1-2\sqrt{1-\mathcal{Y}}\cos{\left(\frac{\theta+2\pi}{3}\right)}\right]\right\}^{1/2},\nb\\
y_2&=&\left\{\frac{b_1}{3b_2 \epsilon_*^2 }\left[1-2\sqrt{1-\mathcal{Y}}\cos{\left(\frac{\theta+4\pi}{3}\right)}\right]\right\}^{1/2},\nb\\
\eqn
with $\mathcal{Y}\equiv 3b_2/b_1^2$ and
\bqn
\theta\equiv - \left(1-\frac{3}{2}\mathcal{Y}+\frac{3}{2}b_1 \mathcal{Y}^2 \nu^2 \epsilon_*^2\right)(1-\mathcal{Y})^{-3/2}.
\eqn
Depending on the signs of $\Delta$, where
\bqn
\Delta\equiv (\mathcal{Y}-1)^{3}+\left(1-\frac{3}{2}\mathcal{Y}+\frac{3}{2}b_1 \mathcal{Y}^2 \nu^2 \epsilon_*^2\right)^2,
\eqn
the nature of the three roots is different. In Fig.~\ref{fig_gofy}, we display five different cases for the function $\lambda^2 \hat g(y)$, corresponding to different types of roots \footnote{It should be noted that in \cite{uniform_nonlinear} Cases (c) and (d) were considered as one, while Case (e) was limited by requiring that the mode be stable at the UV.}. When $\Delta<0$, the three roots $(y_0, y_1, y_2)$ are all real and different,  which corresponds to Case (a) in Fig. \ref{fig_gofy}. When $\Delta=0$, there are one single real root $y_0$ and one double real root $y_1$ ($y_1=y_2$), which corresponds to Case (b) in Fig. \ref{fig_gofy}. When $\Delta>0$, there are one single root $y_0$ and two complex conjugated roots $y_1$ and $y_2$  with $y_1*=y_2$, which corresponds to Cases (c) and (d) in Fig. \ref{fig_gofy}. The difference between Case (c) and Case (d)  is that for Case (d), the two complex roots are largely spaced in the imaginary axis, and as we shall show later, we can treat this case as it has only one single real root. We also display a special Case (e) in Fig. \ref{fig_gofy} by taking $\hat b_2 <0$, which has two real turning points. Note that throughout this paper we assume $y_0<\text{Re}(y_1)\leq \text{Re}(y_2)$.

\begin{figure}
\includegraphics[width=8.1cm]{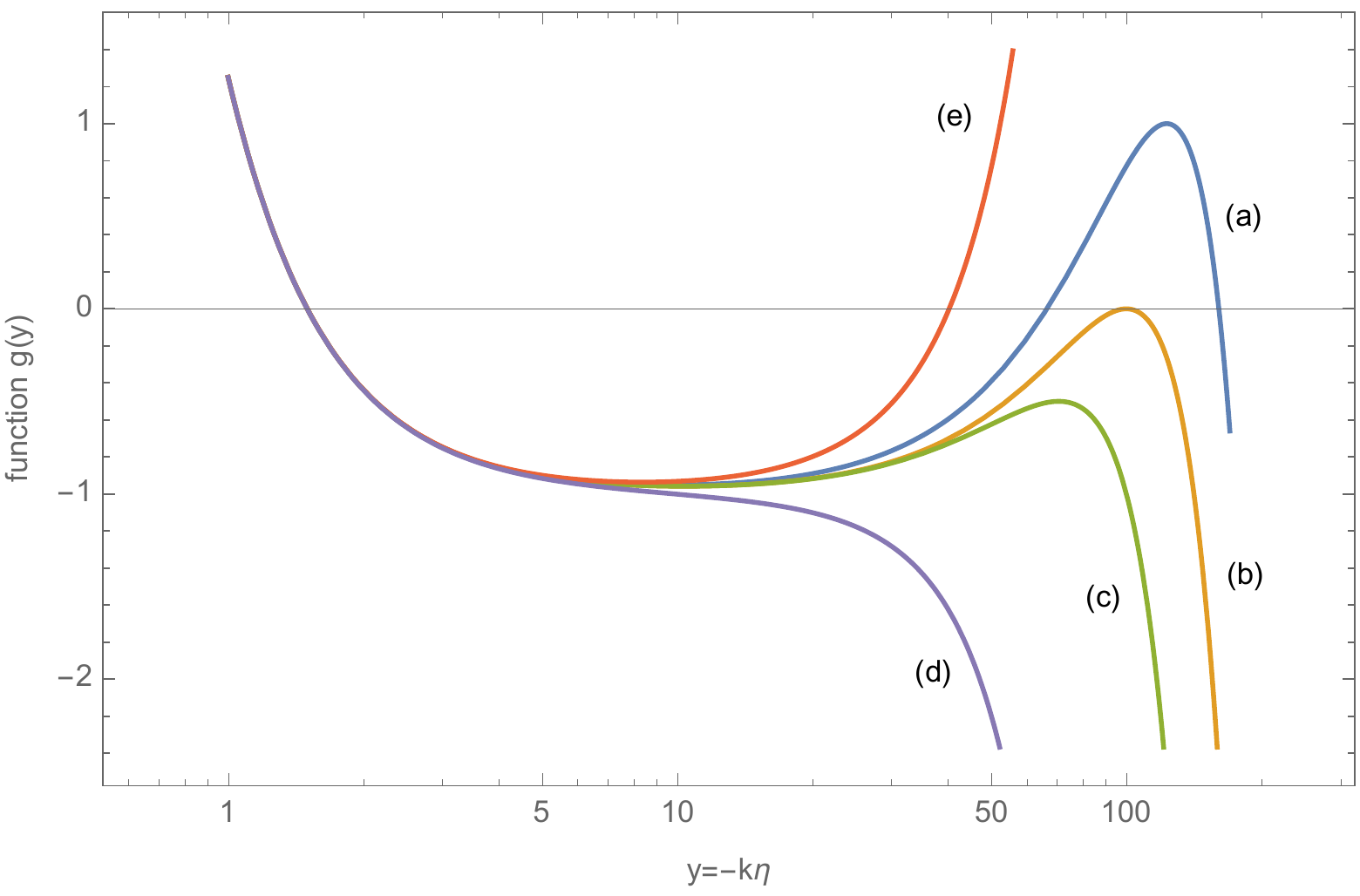}
\caption{The function $g(y)=\lambda^2 \hat g(y)$ defined in Eq.(\ref{ggofy}): (a) Three different and single real roots of the equation $\lambda^2 \hat g(y)=0$. (b) One single real root and one double real root. (c) One single real root and two complex roots. (d) One single real root. (e) Two single real roots when $\hat b_2 <0$. The turning points $y_0$, $y_1$, and $y_2$ are  all   real and positive in Case (a).} \lb{fig_gofy}
\end{figure}

The performance of the uniform asymptotic approximation for different types of turning points are different. The treatment for the single real turning point is presented in Appendix.~\ref{appendixA3}, while the treatment for a pair of turning points which could be either both single, double, or both complex is presented in Appendix.~\ref{appendixA4}. For the single real turning point $y_0$, following the procedure in Appendix.~\ref{appendixA3}, one can construct the approximate solution of Eq.~(\ref{eom}) for primordial perturbation modes as
\bqn\lb{sol_y0}
\mu_k(y) =  \left(\frac{\xi(y)}{\hat g(y)}\right)^{1/4} U(\xi),
\eqn
where $\xi(y)$ is a growing function of $y$ in the range $y \in (0^+, \text{Re}(y_1))$ given by Eq.~(\ref{xiy0}) and $U(\xi)$ is expanded in terms of Airy function in Eq.~(\ref{Uy0}). 

For a pair of turning points $y_1$ and $y_2$, as we discussed above, they could be either both single real, double, or even complex. In the uniform asymptotic approximation, as we have shown in Appendix.~\ref{appendixA4}, we can treat them together. Following the procedure in Appendix.~\ref{appendixA4}, the corresponding solution of Eq.~(\ref{eom}) now can be expressed as
\bqn\lb{sol_y1y2}
\mu_k(y) = \left(\frac{\zeta(y)^2-\zeta_0^2}{-\hat g(y)}\right)^{\frac{1}{4}} U(\zeta),
\eqn
where $\zeta(y)$ is a decaying function of $y$ in the range of $y\in (y_0,+\infty)$, which is given by Eqs. (\ref{zeta1}, \ref{zeta2}, \ref{zeta3}, \ref{zeta4}) respectively for different cases, $\zeta_0^2$ is defined in Eq.~(\ref{zeta_0}),  and $U(\zeta)$ is expressed in terms of parabolic cylinder functions in Eq.~(\ref{Uy1y2}).

Then imposing the adiabatic vacuum state in the limit $y \to +\infty$ as in Eq.~(\ref{ad_vacuum}), and matching the individual solutions in Eq.~(\ref{sol_y0}) and Eq.~(\ref{sol_y1y2}) in their overlaping region $y \in (y_0, \text{Re}(y_1)$, the behaviors of the solution of the primordial perturbations modes during the whole region $y \in (0^+, +\infty)$ now can be completely determined.

As the construction of the approximate solutions in the uniform asymptotic approximation are very much mathematically involved, in this paper we leave  most of the mathematical derivations into Appendix.~\ref{appendixA} and Appendix.~\ref{appendixB}. In Appendix.~\ref{appendixA}, we provide a general and self-contained introduction to the uniform asymptotic approximation and illustrate how to construct accurate  solutions for different types of turning points. In Appendix.~\ref{appendixB}, we show  how to match the individual solutions in different regions so that we can get a solution in the whole spacetime. 

\subsection{Spectra of Primordial Perturbations}

Now with the approximate solutions of Eq.~(\ref{sol_y0}) and Eq.~(\ref{sol_y1y2}), we are in the position to calculate the spectra of  both scalar and tensor perturbations. In order to do so, we  consider the limit $y\to 0^+$, for which only the growing mode of the approximate solution is relevant. Considering the asymptotic form of the Airy functions in the limit $y\to 0^+$ (i.e., $\xi(y)\to +\infty$), we have \cite{uniform_singleH}
\bqn
\lim_{y\to 0^+}\mu_k(\eta)&=&\frac{\beta_0 e^{\frac{2}{3}\xi^{2/3}}}{\lambda^{1/6} \hat g^{1/4} \pi^{1/2}} \Bigg[1+\frac{\mathscr{H}(+\infty)}{2\lambda}\nb\\
&&~~~~+\frac{\mathscr{H}^2(+\infty)}{8\lambda^2}+\mathcal{O}\left(\frac{1}{\lambda^3}\right)\Bigg],
\eqn
where $\mathscr{H}(\xi)$ is the error control function associated with the approximate solution at the turning point $y_0$ and the coefficient $\beta_0$ is determined in Eq.~(\ref{alpha_beta_0}). Then, the power spectra read
\bqn\lb{pw_final}
\Delta^2(k)&\equiv& \frac{k^3}{2\pi^2}\left|\frac{\mu_k(y)}{z(\eta)}\right|^2_{y\to 0^+}\nb\\
&\simeq& \mathscr{A}(k)\frac{k^2}{4\pi^2}\frac{-k\eta}{z^2(\eta) \nu(\eta)}\exp{\left(2\lambda \int_{y}^{y_0} \sqrt{\hat g(y')}dy'\right)}\nb\\
&&\times \left[1+\frac{\mathscr{H}(+\infty)}{2\lambda}+\frac{\mathscr{H}^2(+\infty)}{8\lambda^2}+\mathcal{O}\left(\frac{1}{\lambda^3}\right)\right],\nb\\
\eqn
where $\mathscr{A}(k)$ denotes the modified factor due to the presence of the turning points $y_1$ and $y_2$, which reads
\bqn
\mathscr{A}(k)&=&1+2e^{\pi \lambda \zeta_0^2}+2e^{\pi \lambda \xi_0^2/2}\sqrt{1+e^{\pi \lambda \zeta_0^2}} \nb\\
&&~~~~\times \Bigg\{\cos{2\mathfrak{B}}-\frac{\mathscr{I}(\zeta)+\mathscr{H}(\xi)}{\lambda}\sin{2\mathfrak{B}}\nb\\
&&~~~~~~~~-\frac{\big[\mathscr{I}(\zeta)+\mathscr{H}(\xi)\big]^2}{2\lambda^2}\cos{2\mathfrak{B}}\Bigg\}.\nb\\
\eqn
Here $\mathscr{I}(\zeta)$ is the error control function associated with the approximate solution at the turning points $y_1$ and $y_2$, and the quantity $\mathscr{B}$ is given by Eq.~(\ref{Bcc}).

The quantity $\zeta_0^2$ measures the quantum effects of the nonlinear dispersion relation. When $\zeta_0^2$ is positive and large, which corresponds to the case that both $y_1$ and $y_2$ are real [i.e. Case (a) in Fig. \ref{fig_gofy} with the condition that $y_1$ and $y_2$ are largely spaced], one finds
\bqn
e^{\pi \lambda \zeta_0^2} \gg 1,
\eqn
thus the power spectrum is exponentially enhanced. When $\zeta_0^2=0$, which corresponds to the case that the turning points satisfy $y_1=y_2$, i.e. Case (b) in Fig. \ref{fig_gofy}, one finds
\bqn
\mathscr{A}&=&3+2\sqrt{2}\Bigg\{\cos{2\mathfrak{B}}-\frac{\mathscr{I}(\zeta)+\mathscr{H}(\xi)}{\lambda}\sin{2\mathfrak{B}}\nb\\
&&~~~~~~~~~~~~~~-\frac{\big[\mathscr{I}(\zeta)+\mathscr{H}(\xi)\big]^2}{2\lambda^2}\cos{2\mathfrak{B}}\Bigg\}.  ~~~~~~~
\eqn
For the case $\zeta_0$ is negatively large, which corresponds to the case when the turning points $y_1$ and $y_2$ are complex conjugate and largely spaced in the imaginary axis, that is,  Case (d) in Fig. \ref{fig_gofy},  one has
\bqn
e^{\pi \lambda \zeta^2_0} \ll 1.
\eqn
Thus, the power spectrum reduces to the usual one with only one single turning point $y_0$, which has been calculated in detail in \cite{uniform_singleH} up to the third-order approximation.

With the above expression of the power spectrum,  the corresponding spectral index   can be calculated as
\bqn\lb{indices}
n_{s,t} =\frac{ d\ln \mathscr{A}_{s,t}(k) }{d\ln k}+n_{s,t}^{(one)},
\eqn
where $n_{s,t}^{(one)}$ is the spectral index when $\lambda^2 \hat g(y)$ has only one turning point, and its explicit expression is given  in \cite{uniform_singleH}. When $\zeta_0^2$ is negative and large, $\mathscr{A}_{s,t}$ reduces to one, and thus the first term in  Eq.(\ref{indices}) reduces to zero. This is consistent with the discussions given  in the last paragraph. More specifically, for the first-order approximation the modified factor is only a function of $\zeta_0^2$ and $\mathfrak{B}$, and we can write the spectral index in a more explicit form
\bqn
n_{s,t}=\frac{\partial \ln \mathscr{A}_{s,t}}{\partial \zeta_0^2} \frac{d\zeta_0^2}{d \ln k}+ \frac{\partial \ln \mathscr{A}_{s,t}}{\partial \mathfrak{B}} \frac{d\mathfrak{B}}{d\ln k}+n_{s,t}^{(one)},\nb\\
\eqn
where
\bqn
 \frac{d\zeta_0^2}{d \ln k}\equiv \frac{2}{\pi} \frac{d}{d\ln k} \int_{y_1}^{y_2} \sqrt{\hat g(y')}dy',
\eqn
and
\bqn
 \frac{d\mathfrak{B}}{d\ln k}&\equiv& \lambda \frac{d}{d\ln k} \int_{y_0}^{\text{Re}(y_1)} \sqrt{-\hat g(y')}dy'\nb\\
 &+& \frac{\lambda }{\pi} \phi'\left(\frac{\lambda}{2}\zeta_0^2\right)  \frac{d}{d\ln k} \int_{y_1}^{y_2} \sqrt{\hat g(y')}dy',
\eqn
in which $\phi(x)$ is given by Eq.(\ref{phi}) and $\phi'(x)\equiv d\phi(x)/dx$. Here we would like to mention that the above formulas are very general  and can be applied to any dispersion relation which has three turning points.

Some remarks about the terminologies 
are in order. In general, the function $\lambda^2 \hat g(y)$ in Eq.(\ref{ggofy}) has three turning points $y_0$, $y_1$, and $y_2$, as we already have shown in Fig. \ref{fig_gofy}. In order to apply the uniform asymptotic approximation to the turning point $y_0$, we have imposed the conditions: (i) {\em $|q(y)|$ is small compared with $|\lambda^2 \hat g(y)|$, except in the neighborhoods of $y_0$ where it is small compared with $|\lambda^2 \hat g(y) (y-y_0)^{-1}|$.} {(ii)} {\em At the turning points $y_1$ and $y_2$,   $|q(y)|$ is small  compared with $|\lambda^2 \hat g(y)|$, except in the neighborhoods of $y_1$ and $y_2$ where it is small compared with $|\lambda^2 \hat g(y) (y-y_1)^{-1} (y-y_2)^{-1}|$.} In general the first condition does not hold in the neighborhoods of $y_1$ and $y_2$ [Cases (a), (b) and (c) as shown in Fig. \ref{fig_gofy}], except for the case that $y_1$ and $y_2$ are complex and largely spaced in the complex plane, which is Case (d) as shown in Fig. \ref{fig_gofy}. In the latter, one can consider that the function $\lambda^2 \hat g(y)$ has one turning point in the whole range of $y$, and as we already showed above, the result obtained with three-turning points exactly reduces to the result obtained with one-turning point.  Therefore, when we consider that the function $\lambda^2 \hat g(y)$ only has one-turning point, we mean that  {condition (i)} holds everywhere in the whole range of our interest, which is Case (d) in Fig. \ref{fig_gofy}.

\subsection{Back-reaction of quantum gravitational effects}

As we have mentioned in the introduction, in general the nonlinear dispersion relation leads to the productions of particles. This raises an important question, namely whether or not the back-reaction of the excited modes is small enough to allow inflation to last longer enough. In order to clarify this point, let us consider the approximate solution in the sub-horizon region $H<k/a<M_*$ (equivalently for the region $y_0 \ll y \ll \text{Re}(y_1))$), during which the approximate solution takes the form as Eq.~(\ref{asy1-expansion}). In the sub-horizon region, as $\omega_k^2 \gg z''/z$, thus one has $\lambda^2 \hat g(y) \simeq - \omega_k^2/k^2$ and
\bqn
\lambda \int_{y_0}^y \sqrt{-\hat g}dy' \simeq - \int_{\eta_0}^\eta \omega_k(\eta)d\eta,
\eqn
then the solution in Eq.~(\ref{asy1-expansion}) can be simplified into the form
\bqn
\mu_k(\eta)& \simeq &\frac{\lambda^{1/3}}{\sqrt{2 \omega_k}} \sqrt{\frac{k}{2\pi}} \left(A-i\frac{\bar B}{\lambda}\right)\nb\\
&&\times \Big\{(\alpha_0+i\beta_0) e^{i \int_{\eta_0}^{\eta}\omega_k d\eta'-\frac{\pi}{4}} \nb\\
&&~~~~~+(\alpha_0-i \beta_0) e^{-i \int_{\eta_0}^{\eta}\omega_k d\eta'+\frac{\pi}{4}}\Big\},
\eqn
where $A=A(\lambda,\xi)$, $\bar B(\lambda,\xi)/\lambda$ are given by Eq.~(\ref{ABcc}), and $\alpha_0$, $\beta_0$ are given by Eq.(\ref{alpha_beta_0}). From the above analytical approximate solutions, one can identify the Bogoliubov coefficient of the excited modes at the sub-horizon scales as,
\bqn
\frac{|\tilde \beta_k|^2}{\lambda^{2/3}} = \frac{k}{2\pi}\left|A-i\frac{\bar B}{\lambda}\right|^2 |\alpha_0+i \beta_0|^2.
\eqn
By using Eq.~(\ref{ABcc}) one can obtain that
\bqn
 \left|A-i\frac{\bar B}{\lambda}\right|^2 \simeq 1+\mathcal{O}(1/\lambda^3),
\eqn
thus up to the third-order approximation in the uniform asymptotic approximation, one finds
\bqn
\frac{|\tilde \beta_k|^2}{\lambda^{2/3}} =\frac{k}{2\pi}\Big[|\alpha_0|^2+|\beta_0|^2+i(\alpha^*_0\beta_0-\alpha_0\beta^*_0)\Big].
\eqn
In order to avoid large back-reactions [cf. Eq.(\ref{constriantAa})], one has to impose the condition 
 \cite{back-reaction},
\bqn\lb{constraintA}
|\tilde \beta_k|^2 \lesssim 8 \pi \frac{H_{\text{inf}}^2 M_{\text{Pl}}^2}{M_*^4},
\eqn
where $H_{\text{inf}}$ is the energy scale of the inflation and the Planck 2015 data yields the constraint $H_{\text{inf}}/M_{\text{Pl}} \leq 3.5 \times 10^{-5}$ \cite{PLANCK2015}. Thus if we take $H_{\text{inf}}/M_* \sim 2\times  10^{-3}$, one can infer that
\bqn\lb{constraint_one}
|\tilde \beta_k|^2 \lesssim \mathcal{O}(1).
\eqn
For smaller $M_*$, the Bogoliubov coefficient $ |\tilde \beta_k|^2$ can be larger \footnote{For example, in the healthy extension of the Ho\v{r}ava theory \cite{Horava}, solar system tests lead to
$M_* \lesssim 10^{16} {\mbox{GeV}}$ \cite{BPS}.}. 
But here we shall take the above limit and 
derive the constraints on $\mathscr{A}_{s,t}$. Since the modified factor in general can be expressed as
\bqn
\mathscr{A} \simeq |\tilde \alpha_k+\tilde \beta_k|^2,
\eqn
 it is easy to obtain that
\bqn
\sqrt{1+|\tilde \beta_k|^2}- |\tilde \beta_k|\lesssim \sqrt{\mathscr{A}} \lesssim |\tilde \beta_k|+\sqrt{1+|\tilde \beta_k|^2}.\nb\\
\eqn
Then, Eq.(\ref{constraint_one}) places constraints on the modified factor $\mathscr{A}_{s,t}$ for both of the primordial scalar and tensor perturbations as
\bqn\lb{constraint_on_A}
3-2\sqrt{2} \lesssim \mathscr{A}_{s,t} \lesssim 3+2 \sqrt{2}.
\eqn

\section{Main Features of the Spectra}
\renewcommand{\theequation}{3.\arabic{equation}} \setcounter{equation}{0}

One natural question is whether the quantum gravitational effects could produce some non-trivial features in the spectra of the primordial perturbations. In this section, we shall point out some of these features that could be observationally interesting.

\subsection{Nearly scale-invariance of the primordial perturbations}

We consider the nonlinear dispersion relation with
\bqn
b_1 =\left(1+\mathcal{O}(\epsilon)\right)\hat b_1,\\
b_2 = \left(1+\mathcal{O}(\epsilon)\right)\hat b_2,
\eqn
where $\epsilon$ represents the slow-roll parameter. From these expressions we see that if we expand the perturbations spectra about a pivot scale $k_\star$, the derivative of all the above quantities only contributes to the second-order of the slow-roll approximations or to the order $\mathcal{O}(\epsilon)\times \epsilon_*^2$ for the $\epsilon_*$ term. So if we only consider the first-order slow-roll expansion (ignore the term $\mathcal{O}(\epsilon)\times \epsilon_*^2$ as well), the spectral indices  will be the same as for the case that $\lambda^2 \hat g(y)$ only has one turning point, in which  the quantum gravitational effects  contribute only  small corrections to the spectral index of GR. Thus, even after the quantum gravitational effects are taken into account, in this case the primordial spectra are still nearly scale-invariant.

This is  important. As discussed in the second reference of \cite{Martin2001}, the authors have imposed two additional conditions, the adiabatic condition (i.e., condition when Eq.(\ref{ad_condition}) holds) and the separation of the scale (i.e., $\epsilon_* \ll 1$), to justify the scale invariance of the power spectra. With these two conditions, the evolution of inflationary modes which starts at an initial Bunch-Davies vacuum must trace the adiabatic state during inflation until the mode crosses the Hubble horizon. This exactly corresponds to the one turning point case we discussed above [Case (d) as shown in Fig. \ref{fig_gofy}]. If the adiabatic condition is violated in an intermediate region during inflation, then the equation of the mode function may have more than one turning point [in fact three turning points as shown in Fig. \ref{fig_gofy} for Cases (a), (b), and (c)]. In these cases, the initial Bunch-Davies vacuum shall evolve into a mixed state as shown in Eq.(\ref{wkb_sol}) with a non-zero Bogoliubov coefficient $\tilde \beta_k$ even when the adiabatic condition is restored before these modes leave the Hubble horizon. Our results show that even with such mixed state, the property of almost scale-invariance of the primordial spectrum  still remains.

In addition to the above possibilities, there still exist cases in which   strong scale-dependence of primordial spectrum may occur. For example,  the adiabatic condition may be violated at the initial time. In particular, this is the case  when $\hat b_2 <0$, Case (e) as shown in Fig. \ref{fig_gofy}. In such a case, we cannot choose the initial state as the usual Bunch-Davies vacuum. In the first reference of \cite{Martin2001}, the authors considered the case in which  the initial state is determined by minimizing the energy density. For such an initial state, it was found that   significant deviations from the scale-invariant spectrum can be obtained. However, as we already pointed out in the Introduction, a healthy ultraviolet limit requires $\hat b_2 >0$, thus scale invariance is protected by a stability requirement of the underlying theory in the UV. 

Moreover,  even if a healthy ultraviolet limit is guaranteed, the scale-invariance of the spectrum could still be changed dramatically  when different  initial conditions are chosen.  For example,  it was shown that the power spectrum could be strongly scale-dependent when one chooses an instantaneous Minkowski vacuum, see the first reference of \cite{Martin2001} for details. 



\subsection{Oscillations of the primordial perturbation spectra}

Another important effect from the quantum gravitational corrections is that they generically lead to oscillations in the primordial perturbation spectra \cite{Brandenberger2013CQG}, as one can see clearly from our analytical expression (\ref{pw_final}). Roughly, the phase of the oscillations can be expanded at the pivot scale $k_\star$ in the form,
\bqn
2\mathfrak{B}(k)\simeq 2\mathfrak{B}(k_\star)+\mathcal{O}(\epsilon^2) \times \ln{\frac{k}{k_\star}}.
\eqn
It shows clearly that the second term in the expansion is at the second-order in slow-roll approximation, which indicates that the $k$-dependence is extremely weak, and thus it might be very difficult to observe it in the current and forthcoming experiments. In this sense, the cosine function in Eq.~(\ref{pw_final})  affects only the overall amplitude in primordial spectrum, rather than produces a $k$-dependent oscillatory pattern in the primordial spectrum. It must be noted that the underlying assumption of such a conclusion is that the initial state is the Bunch-Davies vacuum. Similar to the cases studied in the first reference of \cite{Martin2001}, choosing the initial state as an instantaneous Minkowski vacuum could  lead to scale-dependent oscillations in the power spectra that could be observationally significant. But, as we mentioned above, this will also lead to significant derivations of scale-invariance of the power spectra. 


\subsection{Modifications of Power Spectra}

In addition, the  quantum gravitational effects generically modify    the primordial spectra, depending on the coupling constants $\hat b_1$ and $\hat b_2$, where the values of these parameters depend on the types of perturbations, tensor or scalar
\cite{HL}. This will in turn affect the ratio $r$ between the tensor and scalar power spectra, which could lead to observational consequences, and bring the theory directly under tests. For example, in GR the inflationary model with a quadratic potential predicts,
\bqn
r_{\text{GR}} \equiv \frac{8 \Delta_{\text{GR}}^{(t) 2} (k_\star)}{\Delta_{\text{GR}}^{(s) 2}(k_\star)} \sim 0.13,
\eqn
which is obviously in tension with the upper bound obtained recently by Planck 2015, $r_{\mbox{Planck}} \leq 0.11$ at $95\%$ C.L. \cite{PLANCK2015}. However, if we take quantum gravitational effects into account,  the tension could disappear completely, as now we have, 
\bqn
r \equiv \frac{8 \Delta^{(t) 2}(k_\star)}{\Delta^{(s) 2} (k_\star)} = \left(\frac{\mathscr{A}_t}{\mathscr{A}_s}\right) r_{\text{GR}}.
\eqn
Thus, by properly choosing the coupling constants $\hat b^{(t,s)}_A\; (A = 1, 2)$, one has various ways to rescue the quadratic model, for example, by suppressing $\mathscr{A}_t$ and/or enlarging  $\mathscr{A}_s$. As shown in the last section, there is a large room to adjust the value of $\mathscr{A}_t/\mathscr{A}_s$, even after the back-reactions are taken into account, which leads to  the constraints of Eq.(\ref{constraint_on_A}), from which we find that 
\bqn\lb{constraintB}
\left(3-2\sqrt{2}\right)^2  \lesssim \frac{\mathscr{A}_t}{\mathscr{A}_s} \lesssim \left(3+2 \sqrt{2}\right)^2.
\eqn

In Fig. \ref{Fig5} and Fig. \ref{Fig5a}, we display four examples of the modified factor $\mathscr{A}_{n}$ for different sets of parameters $b_1, b_2, \;\epsilon_{*}$. It is shown clearly that the constraint for the modified factor of Eq.(\ref{constraintA}) can be easily realized by properly choosing the values of $b_1$, $b_2$, and $\epsilon_*$. 

It is worthwhile to emphasize that even though with the constraints on $\mathscr{A}_t/\mathscr{A}_s$ of Eq.(\ref{constraintB}), the quantum gravitational effects still provide a viable mechanism   to reconcile the tension between some inflation models and the Planck2015 observational  data. To illustrate this clearly, let us take a look at the simplest inflation model with a power-law potential $V(\phi) =\lambda_{p} \phi^{p}$ as an example. When the quantum gravitational effects are taken into account, as shown in Fig. \ref{Fig6}, in which we  take $\mathscr{A}_t/\mathscr{A}_s$ to be in the range $(0.48, 1.2)$, which is well within the constraints of Eq.(\ref{constraint_on_A}),  the inflation model with a quadratic potential now can be perfectly consistent with Planck 2015 data.

\begin{figure*}
{\label{AmodifiedA}
\includegraphics[width=8.1cm]{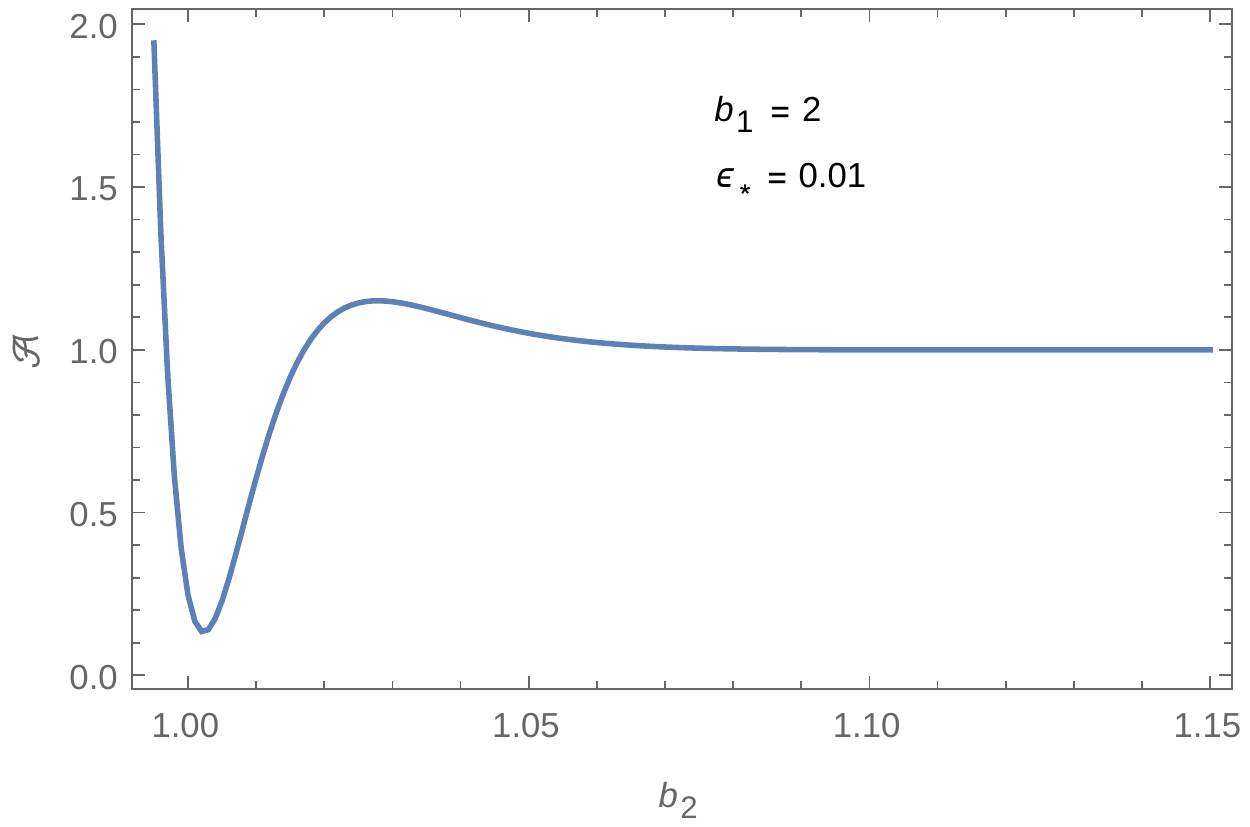}}
{\label{AmodifiedB}
\includegraphics[width=8.1cm]{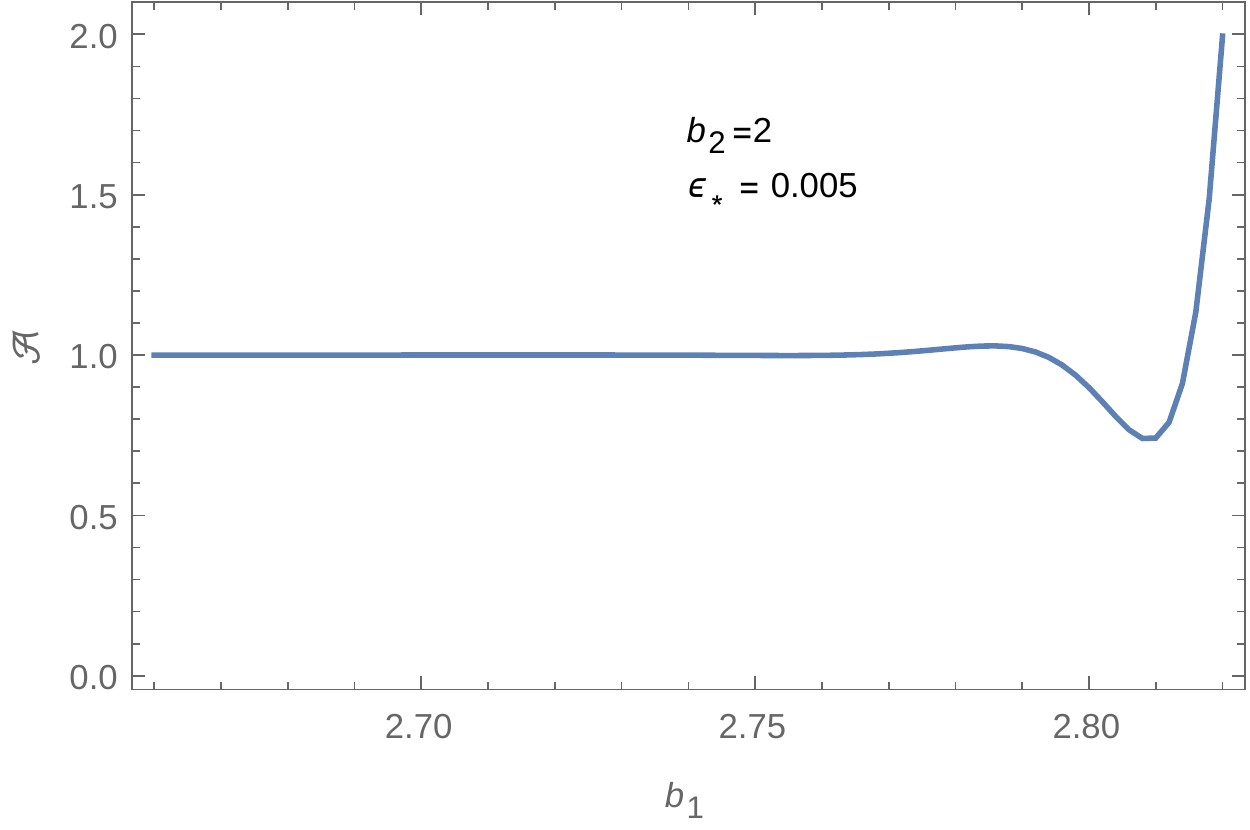}}\\
\caption{The modified factor $\mathscr{A}$ for two different sets of parameters with $\nu=3/2$. Left panel: the modified factor $\mathscr{A}$ vs $b_2$ with $b_1$ and $\epsilon_*$ fixed. Right panel: the modified factor $\mathscr{A}$ vs  $b_1$ with $b_2$ and $\epsilon_*$ fixed.}\lb{Fig5}
\end{figure*}

\begin{figure*}
{\label{Aepsilon1}
\includegraphics[width=8.1cm]{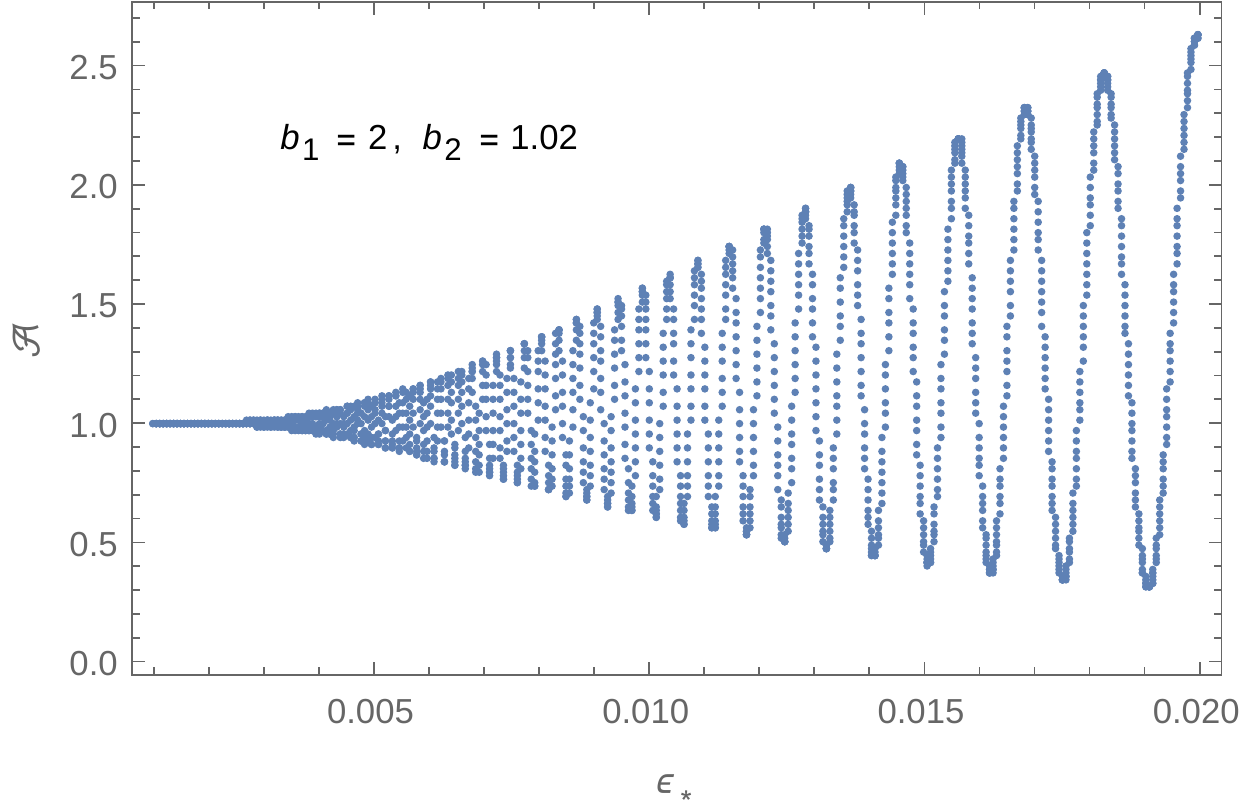}}
{\label{Aepsilon2}
\includegraphics[width=8.1cm]{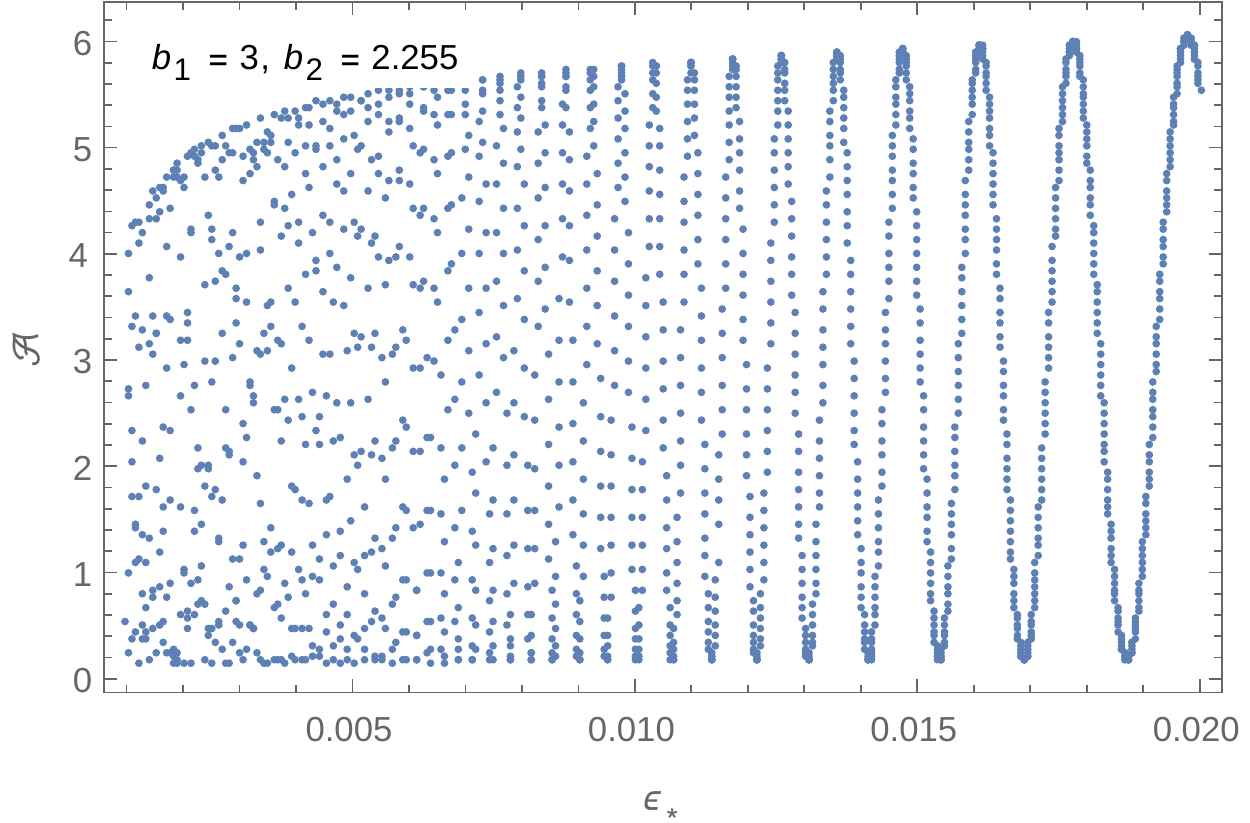}}
\caption{The dependence of the modified factor $\mathscr{A}$ on $\epsilon_*$ for two different sets of parameters $(b_1, b_2)$ but all with $\nu=3/2$. (i) Left panel: the modified factor $\mathscr{A}$ vs $\epsilon_*$ with $b_1=2$ and $b_2=1.02$. In this case, for the range of $\epsilon_*$ we considered in the figure, the function $\lambda^2 \hat g(y)$ has one single real and two complex turning points, which corresponds to Cases (c) and (d) in Fig. \ref{fig_gofy}. When $\epsilon_* $ approaches to zero, the function $\lambda^2 \hat g(y)$ gradually changes from Case (c) to Case (d), and the factor $\mathscr{A}$ approaches to one, as expected. (ii) Right panel: the modified factor $\mathscr{A}$ vs  $\epsilon_*$ with $b_1=3$ and $b_2=2.255$. The values of the parameters chosen in this case  correspond to the case in which the turning points $y_1$ and $y_2$ are real and very close to each other, i.e., $|y_1/y_2-1| \ll 1$, so that  the modified factor $\mathscr{A} \simeq 3+2 \sqrt{2} \cos{2\mathfrak{B}}$.}\lb{Fig5a}
\end{figure*}

\begin{figure*}
{\label{NSRPLOT}
\includegraphics[width=12.1cm]{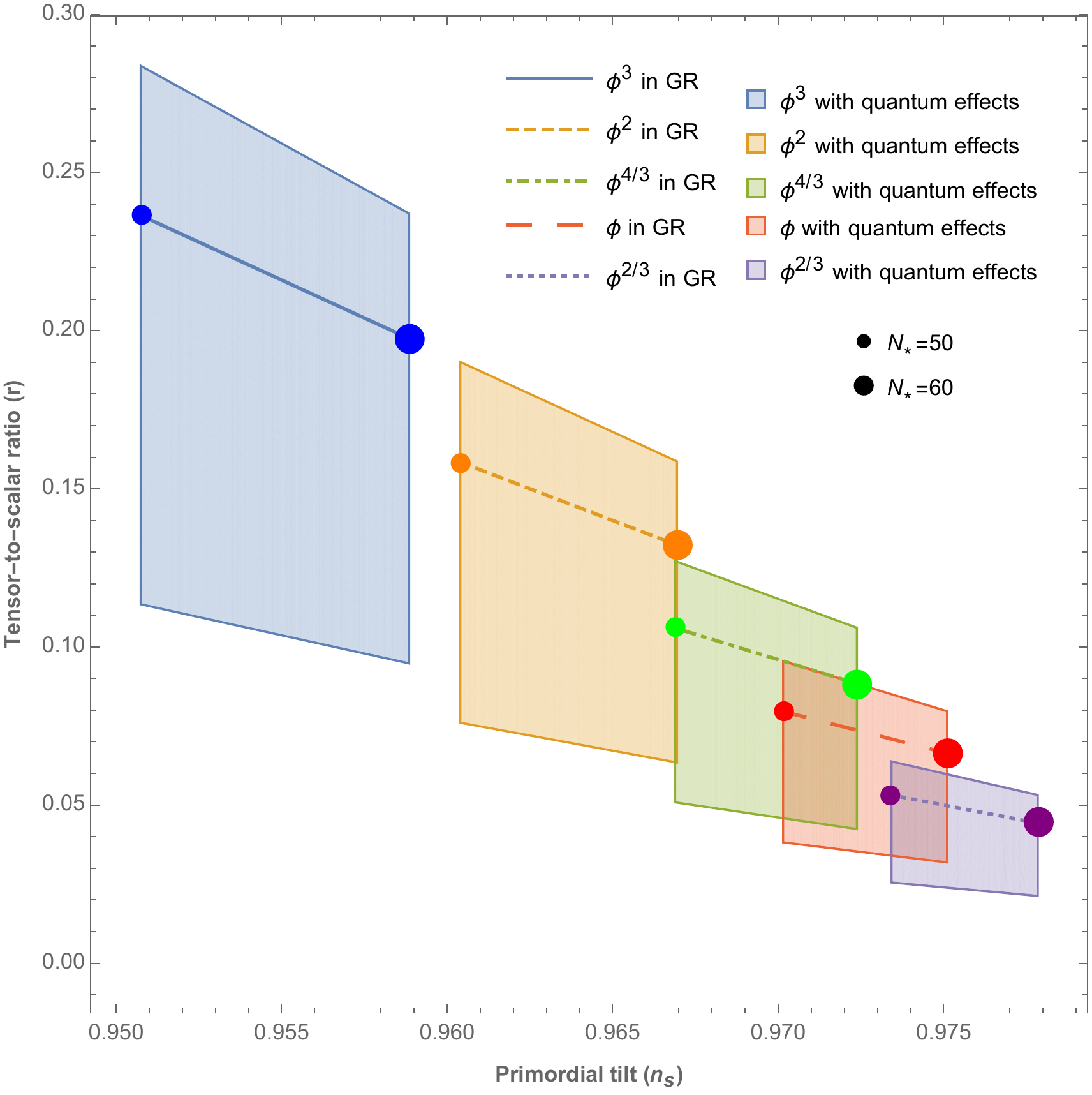}}
\caption{Theoretical predictions for the inflationary models with the potential  $V(\phi) = \lambda_{p} \phi^{p}$ after quantum gravitational effects are taken into account.  In this figure, we take the ratio $\mathscr{A}_t/\mathscr{A}_s$ to be in the range $(0.48,1.2)$, which is well within the constraints  Eq.(\ref{constraint_on_A}), obtained after the back-reactions are taken into account \cite{backreaction,back-reaction,back-reaction2}.  The upper limit of Planck2015 data is $r \le 0.11$ at $95\%$ C.L.
The shaded quadrangle regions are the theoretical allowed regions of the parameters $r$ and $n_s$ for a given $p$. In the case $p = 2$ one can see that $r$ can be as small as $0.07$. As mentioned in the content, it can be even smaller 
for different choices of $\epsilon_*$ \cite{BPS}.}\lb{Fig6}
\end{figure*}

\section{Conclusions}

The uniform asymptotic approximation method provides a powerful, systematically improvable, and error controlled approach to construct analytical solutions of inflationary perturbations. It has been   applied to inflationary models in various situations, in which the traditional methods, such as WKB and Green functions, either are not applicable or produce large errors \cite{JM09}.  These include models with nonlinear dispersion relations \cite{uniform_nonlinear, uniform_singleH}, $k$-inflation \cite{uniformK}, and holonomy and inverse-volume corrections from LQC \cite{uniform_loop}.

In this paper, we have systematically generalized  our previous studies of quantum gravitational effects with the nonlinear dispersion relation (\ref{nonlinear}) from the first-order approximations \cite{uniform_nonlinear} to the third-order with any number and  order of turning-points.  The upper bounds of errors to the third-order approximations  are $\le 0.15\%$ \cite{uniform_singleH}, which are  accurate enough   for the current and forthcoming experiments \cite{S4-CMB}.

To our goals, we have first derived  the analytic approximate solutions of perturbations associated with different turning points to any order of approximations, and then matched them together to the third-order. With the matched solutions, we have been able to calculate explicitly the primordial tensor and scalar perturbation spectra, which represents the most accurate results in the literature, as far as we know. From these expressions, we have also derived several   features of the power spectra due to quantum gravitational effects that may be observationally interesting.  

After deriving explicitly the power spectra of the tensor and scalar perturbations up to the third-order approximation, we have investigated the constraints due to the back-reactions of quantum gravitational effects \cite{backreaction,back-reaction,back-reaction2}. This is very important,  in order to make sure that these effects will not make inflation end in a very early period, and  the problems, such as horizon, flatness and monopole, are still solved. With a very conservative assumption about the energy scale $M_*$ of the underlying theory of quantum gravity, we have found the bounds on the amplitudes of the power spectra, which are given explicitly by Eq.(\ref{constraintB}). Even with these severe constraints, we have found the tensions between certain inflationary models and observations can be reconciled easily, after quantum gravitational effects are taken into account. These include the chaotic model with a quadratic potential.

Finally, we would like to note that, although we have studied the quantum gravitational effects specified by the nonlinear dispersion relation (\ref{nonlinear}), in which the equation of motion could have three turning points, we would like to emphasize that the method and results presented in this paper  can be easily extended to other cases. In particular, the method is equally well applicable to non-slow-roll inflationary models. 

\section*{Acknowledgements}

 Part of the work was    carried out when A.W. was visiting the State University of Rio de Janeiro (UERJ), Brazil. A.W. would like to express his gratitude to UERJ for hospitality.    
This work is supported in part by Ci\^encia Sem Fronteiras, No. 004/2013 - DRI/CAPES, Brazil (A.W.); Chinese NSFC No. 11375153 (A.W.), No. 11173021 (A.W.), 
  No. 11047008 (T.Z.), No. 11105120 (T.Z.), and No. 11205133 (T.Z.).
  
 
 \appendix
 
\section{\label{appendixA}The uniform asymptotic approximation}
\renewcommand{\theequation}{A.\arabic{equation}}
\setcounter{equation}{0}
In this appendix, we provide the novel ideas of the uniform asymptotic approximation with high-order approximations. The approximation scheme presented here provides a systematic and powerful approach to the second-order ordinary differential equation
with  various turning points, at which the WKB approximation becomes invalid. To make our formulas as much applicable as possible, in this appendix we shall present the development in a very general manner, so that we could easily extend the   results 
 to other circumstances. 

\subsection{\lb{appendixA1}Equation of the Mode Function and Liouville Transformation}

Our starting point is to consider the following second-order differential equation \cite{Olver1974, Olver1975},
\bqn\lb{standarda}
\frac{d^2\mu_k(y)}{dy^2}=\Big[\lambda^2 \hat{g}(y)+q(y)\Big]\mu_k(y),
\eqn
where $\lambda^2$ is supposed to be a large positive parameter, and the functions $\lambda^2 \hat g(y)$ and $q(y)$ will be determined by the analysis of the error bounds given below, so that the associated errors will be minimized. In general, $\lambda^2 \hat g(y)$ and $q(y)$ could have poles or zeros in the interval of our interest. We also called the zeros of $\lambda^2 \hat g(y)$ as turning points of the equation (\ref{standarda}). From the theory of the second-order differential equation, the uniform asymptotic solution of $\mu_k(y)$ depends on the behavior of the functions $\lambda^2 \hat g(y)$ and $q(y)$ around their poles (singularities) and zeros (turning points). Note that in this paper, we also use the notation $g(y)=\lambda^2 \hat g(y)$, and in practice, when we turn to the final results, we may set $\lambda=1$ for simplification.

To proceed further, let us introduce the Liouville transformation with two new variable $U(\xi)$ and $\xi(y)$ \cite{Olver1974, Olver1975},
\bqn\lb{Olver trans}
U(\xi)=\chi^{1/4} \mu_k(y), \;\;\;\chi=\xi'^2=\frac{|\hat{g}(y)|}{f^{(1)}(\xi)^2},
\eqn
where $\xi'=d\xi/dy$, and
\bqn
f(\xi)=\int \sqrt{|\hat{g}(y)|}dy,\;\;\;f^{(1)}(\xi)=\frac{df(\xi)}{d\xi}.
\eqn
Note that $\chi$ must be regular and not vanish in the intervals of interest. Consequently, the function $f(\xi)$ must be chosen so that $f^{(1)}(\xi)$ has zeros and singularities of the same type as those of $\hat g(y)$. As shown below, such requirements play an essential role in determining the approximate solutions. In terms of $U(\xi)$ and $\xi(y)$, Eq.(\ref{standard}) is brought into the form,
\bqn\lb{eomU}
\frac{d^2 U(\xi)}{d\xi^2} =\Big[\pm \lambda^2 f^{(1)}(\xi)^2+\psi(\xi)\Big]U(\xi),
\eqn
where
\bqn
\psi(\xi)&=&\frac{q(y)}{\chi}-\chi^{-3/4} \frac{d^2 (\chi^{-1/4})}{dy^2}.
\eqn
Here $\pm$ correspond to $\hat{g}(y)>0$ and $\hat{g}(y)<0$, respectively, and $\chi'\equiv d\chi/dy$. Considering $\psi(\xi)=0$ as the first-order approximation, one must choose $f^{(1)}(\xi)^2$ so that: (a) the first-order approximation is as close to the exact solution as possible, 
and (b) the resulting equation can be solved explicitly (in terms of some known special functions). Clearly, such a choice sensitively depends on the behavior of the functions $\lambda^2 \hat g(y)$ and $q(y)$ near the poles and zeros. Therefore, in the following let us first consider solutions near poles.


\subsection{\lb{appendixA2}Liouville-Green approximations near poles}

\subsubsection{\lb{appendixA2a}Liouville-Green approximate solutions and their error bounds}

In most of physical systems, the regions of physical interest are in general in the intervals between two poles or on one side of them. In this section, let us assume that the functions $\lambda^2 \hat g(y)$ and $q(y)$ have two poles, one is located at $y= 0^{+}$ and another is at $y= +\infty$. In addition, we assume $\hat g(y)>0$ near $y=0^+$ and $\hat g(y)<0$ when $y\to +\infty$. Except the regions near the zeros of $\lambda^2 \hat g(y)$, the functions $\lambda^2 \hat g(y)$ and $q(y)$ are usually well-defined in their neighborhoods of poles. With this property, we can choose
\bqn\lb{fpole}
f^{(1)}(\xi)^2=\text{const}.
\eqn
Without loss of generality, we can always set  this constant to one. Then, from Eq.~(\ref{Olver trans}) we find
\bqn
\xi(y)=\int \sqrt{\pm \hat g(y)} dy,
\eqn
here ``+" (``-") corresponds to $\hat g(y)>0$ and $\hat g(y)<0$ respectively, and the equation of motion (\ref{eomU})  takes the form
\bqn\lb{eompole}
\frac{d^2U(\xi)}{d\xi^2}=\left[\pm \lambda^2+\psi(\xi)\right] U(\xi).
\eqn

Let us first consider the approximate solution near the pole $ y = 0^{+}$. In this case, we choose $\xi(y)$ as a monotone increasing function of $y$, thus we find
\bqn
\xi(y)=\int^y_{y_i} \sqrt{\hat g(y')}dy',
\eqn
where $y_i$ is an irrelevant reference point near $y=0^+$. Then, the equation of motion now reads
\bqn\lb{EoM0}
\frac{d^2U(\xi)}{d\xi^2}=\left[\lambda^2+\psi(\xi)\right] U(\xi).
\eqn
To the first-order approximation, that is, neglecting the $\psi(\xi)$ term in Eq.~(\ref{EoM0}), the approximate solution takes the form $e^{\pm \lambda \xi}$. Then,  the solution of Eq.~(\ref{EoM0}) with error terms can be constructed as
\bqn\lb{Uplus}
U^{+}=c_{+} e^{\lambda \xi} (1+\epsilon^{+}_1)+d_{+}e^{-\lambda \xi} (1+\epsilon^{+}_2),
\eqn
where $c_{+}$, $d_{+}$ are integration constants, and $\epsilon^{+}_1$ and $\epsilon^{+}_2$ represent the errors of the approximate solution, which are bounded by
\bqn
|\epsilon_1^{+}|,\;\;\;\frac{|d\epsilon_1^{+}/dy|}{2 |\hat g|^{1/2}} \leq \exp{\left(\frac{1}{2} \mathscr{V}_{0,y}(\mathscr{F})\right)}-1,\nb\\
|\epsilon_2^{+}|, \frac{\left|d\epsilon_2^{+}/dy\right|}{2|\hat g|^{1/2}} \;\leq \exp{\left(\frac{1}{2}\mathscr{V}_{y,+\infty}(\mathscr{F})\right)}-1.
\eqn
The derivation of the error bounds can be found in \cite{Olver1974, uniform_nonlinear}. Here $\mathscr{F}(y)$ represents the associated error control function of the approximate solution, which is given by
\bqn\lb{error infty}
\mathscr{F}(y)&=&\int |\psi(v)|dv\nb\\
&=&\int \left[\frac{1}{|g|^{1/4}}\frac{d^2}{dy^2}\left(\frac{1}{|g|^{1/4}}\right)-\frac{q}{|g|^{1/2}}\right] dy, ~~~~~
\eqn
and $\mathscr{V}_{x_1,x_2}(\mathscr{F})$ is the total variation of the function $\mathscr{F}(y)$. Accordingly, the mode function $\mu_k(y)$ is given by  the Liouville-Green (LG) solution
\bqn\lb{LG1}
\mu^{+}_k(y)&=& \frac{c_{+}}{\hat g(y)^{1/4}} e^{\lambda \int^y_{y_i} \sqrt{\hat g(y')}dy'} (1+\epsilon^{+}_1) \nb\\
&&\;\;\;+ \frac{d_{+}}{\hat g(y)^{1/4}} e^{-\lambda \int^y_{y_i} \sqrt{\hat g(y')}dy'} (1+\epsilon^{+}_2).\nb\\
\eqn

Similarly, near the pole $ y = +\infty$, we still choose $\xi(y)$ as a monotone increasing function of $y$. So we have
\bqn
\xi(y)=\int_{y_e}^{y}\sqrt{-\hat g(y')}dy',
\eqn
where $y_e$ is an irrelevant reference point near $y=+\infty$. Then Eq.~(\ref{eompole}) has the form,
\bqn\lb{EoMinf}
\frac{d^2U(\xi)}{d\xi^2}=\left[-\lambda^2+\psi(\xi)\right] U(\xi).
\eqn
Ignoring the term $\psi(\xi)$ as the first-order approximation, the solution of the above equation now takes the form  $\sim e^{\pm i \lambda \xi}$. Thus, the solution of Eq.~(\ref{EoM0}) with the error terms near the pole $y\to \infty$ can be constructed as
\bqn\lb{Uminus}
U^{-}= c_{-} e^{i \lambda \xi }(1+\epsilon_1^{-})+d_{-} e^{-i \lambda \xi} (1+\epsilon_2^{-}),
\eqn
where $c_{-}$, $d_{-}$ are other two integration constants, and $\epsilon^{-}_1$ and $\epsilon^{-}_2$ represent the associated errors of the approximation, and are bounded by
\bqn
|\epsilon_{1}^{-}|,\;\frac{|d\epsilon_1^{-}/dy|}{|g|^{1/2}} \leq \exp{\left(\mathscr{V}_{y,+\infty}(\mathscr{F})\right)}-1,\lb{errorbound1minus}\\
|\epsilon_{2}^{-}|,\;\frac{|d\epsilon_2^{-}/dy|}{|g|^{1/2}} \leq \exp{\left(\mathscr{V}_{0,y}(\mathscr{F})\right)}-1.\lb{errorbound2minus}
\eqn
Then the mode function $\mu^{-}_k(y)$ takes the form
\bqn\lb{LG2}
\mu^{-}_k(y)&=&\frac{c_{-}}{(-\hat g(y))^{1/4}} e^{i \lambda \int^y_{y_e} \sqrt{-\hat g(y')}dy'} (1+\epsilon^{-}_1)\nb\\
&&+\frac{d_{-}}{(-\hat g(y))^{1/4}} e^{-i \lambda \int^y_{y_e} \sqrt{-\hat g(y')}dy'} (1+\epsilon^{-}_2).\nb\\
\eqn

\subsubsection{\lb{appendixA2b}Convergence of error control function near poles }

The LG approximation is meaningful only when the following two conditions hold. First, in the interval of interest, one has to require that $|q(y)|$ is smaller than $\lambda^2 \hat g(y)$ near both of the two poles. The second condition states that the associated error control function $\mathscr{F}(y)$ must be convergent near these poles. These two conditions provide  guidelines for how to determine the splitting of $\lambda^2 \hat g(y)+q(y)$. 


Let us first consider the region near the pole $y=0^+$, in which the function $\lambda^2 \hat g(y)$ and $q(y)$ are expanded in   the form
\bqn
\lb{expand function}
\lambda^2 \hat g(y)=\frac{1}{y^i}\sum_{s=0}^\infty g_s y^{s},\;q(y)=\frac{1}{y^j} \sum_{s=0}^\infty q_s y^s.
\eqn
Note that in writing out the above expressions we have assumed that $\lambda^2 \hat g(y)$ has a pole of order $i$ and $q(y)$ has a pole of order $j$ at $y=0^+$. Substituting  the above expansions into the integrand of the error control function in Eq.~(\ref{error infty}),  we find
\bqn
&&\frac{1}{\hat{g}^{1/4}}\frac{d^2}{dy^2}\left(\frac{1}{\hat{g}^{1/4}}\right)-\frac{q}{\hat{g}^{1/2}}\nb\\
&&~~~~~~~~~~=\frac{5}{16}\frac{\hat{g}'^2}{\hat{g}^{5/2}}-\frac{1}{4}\frac{\hat{g}''}{\hat{g}^{3/2}}-\frac{q}{\hat{g}^{1/2}},
\eqn
where
\bqn
\frac{5}{16}\frac{\hat{g}'^2}{\hat{g}^{5/2}} \simeq \frac{5}{16} y^{\frac{i}{2}-2}\Big[4g_0^{-1/2}+\mathcal{O}(y)\Big],\nb\\
-\frac{1}{4}\frac{\hat{g}''}{\hat{g}^{3/2}}\simeq -\frac{1}{4}y^{\frac{i}{2}-2}\Big[6g_0^{-1/2}+\mathcal{O}(y)\Big],\nb\\
-\frac{q}{\hat{g}^{1/2}}\simeq -y^{\frac{i}{2}-j}\Big[q_0 g_0^{-1/2}+\mathcal{O}(y)\Big].
\eqn
With the above expansions, now let us discuss the convergence of the error control function $\mathscr{F}(y)$ by considering the following different cases.
\begin{itemize}
\item $i>2$. In this case, the error control function $\mathscr{F}(y)$ is convergent only if $j<\frac{i}{2}+1$, i.e.,
\bqn
i>2,\;\;\;j<\frac{i}{2}+1.
\eqn
\item $i=2$. In this case, one also requires $j=2$ and the error control function $\mathscr{F}(y)$ reads
\bqn
&&\frac{5}{16}\frac{\hat{g}'^2}{\hat{g}^{5/2}}-\frac{1}{4}\frac{\hat{g}''}{\hat{g}^{3/2}}-\frac{q}{\hat{g}^{1/2}}\nb\\
&&~~~~\sim -g_0^{1/2}\left(\frac{1}{4}+q_0\right)y^{-1}+\mathcal{O}(y^{0}),
\eqn
which yields
\bqn
i=2,\;\;j=2,\;\;q_0=-\frac{1}{4}.
\eqn
\item $i<2$. In this case, the error control function $\mathscr{F}(y)$ can not be convergent for any choices of $g_s$ and $q_s$.
\end{itemize}


Let us now turn to the pole $y = +\infty$. Assuming that $\hat g(y)$ and $q(y)$ have a pole of order $i$  and $j$, respectively, we find that they can be expanded in the form,
\bqn
\lb{function expand}
g(y)=y^{i} \sum_{s=0}^\infty \bar{g}_s y^{-s},\; q(y)=y^{j} \sum_{s=0}^\infty \bar{q}_s y^{-s},
\eqn
where $\bar{g}_s$ and $\bar{q}_s$ are other sets of constants.   Substituting the above expansions into the integrand of the error control function $\mathscr{F}(y)$ in Eq.~(\ref{error infty}), we find that
\bqn
&&\frac{1}{|g|^{1/4}} \frac{d^2}{dy^2} \left(\frac{1}{|g|^{1/4}}\right)-\frac{q}{|g|^{1/2}} = y^{-2-i/2} \sum_{s=0}^{+\infty} c^{(1)}_s y^{-s}\nb\\
&&~~~~~~~~~~~~ +y^{j-i/2} \sum_{s=0}^{+\infty} c_s^{(2)} y^{-s},
\eqn
where the coefficients $c_s^{(1)}$ and $c_s^{(2)}$ are functions of $\bar{q}_s$ and $\bar{g}_s$. Then,  the convergence of  $\mathscr{F}(y)$ requires
\bqn
\lb{ConditionB}
i>-2, \;\;\;j<\frac{i}{2}-1.
\eqn

The validity of the uniform asymptotic approximation is very sensitive to the types of the poles of the function $\lambda^2 \hat g(y)$. For the sake of further applications of the approximations, let us summarize the main conditions of the validity of the uniform asymptotic approximation as follows:
\begin{itemize}
\item $|q(y)| < |\lambda^2 \hat g(y)|$ near both poles;
\item at the  pole $y=0^+$,  the function $\lambda^2 \hat g (y)$ must be of order $i \geq 2$;
\begin{itemize}
\item for $i>2$, one has to choose $j<\frac{i}{2}+1$; 
\item for $i=2$, one has to choose $q_0=-1/4$ and $j=2$; 
\end{itemize}
\item at the  pole $y=+\infty$, the functions $\lambda^2 \hat g(y)$ and $q(y)$ must chosen so that $i>-2$ and $j<i/2-1$.
\end{itemize}
Note that in the above  $i$ denotes the order of poles of the function $\lambda^2 \hat g(y)$ and $j$ denotes the order of the poles of the function $q(y)$. The physically interesting region lies in the range $y\in (0,+\infty)$, which may contain various turning points. Now, let us turn to the approximation solutions near turning points.

\subsection{\lb{appendixA3}Approximate solution near the single turning point $y_0$}


We first consider the single turning point at $y=y_0$ for the function $\lambda^2 \hat g(y)$. Then,  the function $\hat g(y)$ can be written in  the form
\bqn\lb{g_py0}
\hat g(y)=p(y) (y-y_0),
\eqn
where $p(y)$ is a regular function  with $p\left(y_0\right) \not= 0$ and we also assume $p(y)<0$ \footnote{This assumption is set to be consistent with case we are going to apply in this paper.}. 
In order to perform the uniform asymptotic approximation, we assume that  
 {\em $|q(y)|$ is small in comparison  with $|\lambda^2 \hat g(y)|$, except in the neighborhood of $y_0$, in  which  it must be smaller than  $\left|\lambda^2 \hat g(y) (y-y_0)^{-1}\right|$} \cite{uniform_nonlinear}. As we already mentioned previously, one has to choose $f^{(1)}(\xi)^2$ so that it has the same type of zeros or singularities as $\hat g(y)$. Thus,  around the single turning point $y_0$, one can introduce a monotonically increasing or decreasing function $\xi$ via the relation,
\bqn\lb{f1}
f^{(1)}(\xi)^2=\pm\xi,
\eqn
where $\xi(y_0)=0$, and $\pm$ correspond to $\xi(y)>0$ and $\xi(y)<0$, respectively. Without loss of the generality, we can choose $\xi$ to have the same sign as $\hat g(y)$, and thus $\xi$ is a monotone decreasing function around $y = y_0$. Combining Eqs.(\ref{Olver trans}) and (\ref{f1}), we find
\bqn
\lb{xiy0}
\xi=\begin{cases}
                   - \left(\frac{3}{2} \int^y_{y_0} \sqrt{-\hat g(y')} dy'\right)^{2/3},& y\geq y_0,\\
                   \left(-\frac{3}{2} \int^y_{y_0} \sqrt{\hat g(y')} dy'\right)^{2/3},& y \leq y_0,
\end{cases}
\eqn
and that Eq.(\ref{eomU}) now reads
\bqn
\lb{eomy0}
\frac{d^2U}{d\xi^2}=\Big(\lambda^2 \xi+\psi(\xi)\Big) U.
\eqn
Neglecting the $\psi(\xi)$ term, we find that the above equation has the approximate analytical solution in terms of Airy type functions
\bqn
U(\xi) \sim \text{Ai}(\lambda^{2/3}\xi)~~~ \text{or}~~~\text{Bi}(\lambda^{2/3}\xi).
\eqn
Then,  the solution of Eq.~(\ref{eomy0}) can be casted into the form
\bqn
U(\xi)=\alpha_0 \Big(\text{Ai}(\lambda^{2/3}\xi) +\epsilon^{(1)}_3\Big)+\beta_0 \Big(\text{Bi}(\lambda^{2/3}\xi)+\epsilon^{(1)}_4 \Big),\nb\\
\eqn
where $\text{Ai}(\xi)$ and $\text{Bi}(\xi)$ are the Airy functions, $\alpha_0$ and $\beta_0$ are two integration constants, and $\epsilon_3^{(1)}$ and $\epsilon_4^{(1)}$ denote the errors of the first-order approximation, which are bounded by \cite{uniform_nonlinear, Olver1974}
\bqn\lb{error3}
&&\frac{|\epsilon_3|}{M(\xi)},\;\frac{|\partial \epsilon_3/\partial \xi|}{ N(\xi)} \leq \frac{ E^{-1}(\xi)}{\lambda_0} \Big\{\exp{\Big(\lambda \mathscr{V}_{\xi,a_3}(\mathscr{H})\Big)}-1\Big\},\nb\\
&&\frac{|\epsilon_4|}{M(\xi)},\;\frac{|\partial \epsilon_4/\partial \xi|}{ N(\xi)} \leq \frac{ E(\xi)}{\lambda_0} \Big\{\exp{\Big(\lambda \mathscr{V}_{a_4,\xi}(\mathscr{H})\Big)}-1\Big\},\nb\\
\eqn
where the auxiliary functions $M(\xi)$, $E(\xi)$, and constant $\lambda_0$ are given in \cite{uniform_nonlinear}. $\mathscr{V}_{x_1,x_2}(F)$ is the supremum of the function $F(x)$ in the range $(x_1,x_2)$. $a_3$ and $a_4$ are, respectively,  the upper and lower limit of the variable $\xi$. And $\mathscr{H}$ is the associated error control function, which is defined as
\bqn
\mathscr{H}(\xi)=\int^{\xi}_0 |v|^{-1/2} \psi(v) dv.
\eqn
Then,  the mode function is given by,
\bqn\lb{airy solution1}
\mu_k(y)&=& \alpha_0  \left(\frac{\xi(y)}{\hat g(y)}\right)^{1/4}\Big(\text{Ai}(\lambda^{2/3}\xi) +\epsilon_3^{(1)}\Big)\nb\\
&&+\beta_0 \left(\frac{\xi(y)}{\hat g(y)}\right)^{1/4}\Big( \text{Bi}(\lambda^{2/3}\xi) +\epsilon_4^{(1)}\Big).\nb\\
\eqn

Now let us turn to extend the above first-order approximation to high orders. For this purpose, following Olver  \cite{Olver1974, uniform_singleH}, we assume that the exact solution of $U(\xi)$ takes the form (for the branches of $\text{Ai}(\lambda^{2/3}\xi)$),
\bqn
&&U(\lambda,\xi)=A(\lambda,\xi)\text{Ai}(\lambda^{2/3}\xi)+\lambda^{-4/3}B(\lambda,\xi) \text{Ai}'(\lambda^{2/3}\xi),\nb\\
\eqn
where for the first-order approximation we have $A(\lambda,\xi)=1$ and $B(\lambda,\xi)=0$. Substituting the above expression into the equation of motion (\ref{eomy0}) and equating the coefficients of $\text{Ai}(x)$ and $\text{Ai}'(x)$, we have
\bqn
2 \lambda^2 A'+B''- \psi B=0,\nb\\
A''+ B+2  \xi B'- \psi A=0.
\eqn
These equations are satisfied by the formal expansions of the forms,
\bqn
A(\lambda,\xi)&=&\sum_{s=0}^{+\infty} \lambda^{-2s } A_s(\xi),\nb\\
B(\lambda,\xi)&=& \sum_{s=0}^{+\infty} \lambda^{-2s} B_s(\xi),
\eqn
so that 
\bqn
2  A'_{s+1}+B''_s- \psi B_s=0,\nb\\
A''_s+ B_s+2  \xi B'_s- \psi A_s=0.
\eqn
Integration of the above two equations yields,
\bqn\lb{AB}
&& A_0(\xi)=1,\;\;\nb\\
&&B_s=\frac{\pm 1}{2 (\pm \xi)^{1/2}}\int_0^\xi \{\psi(v) A_s(v)-A''_s(v)\}\frac{dv}{(\pm v)^{1/2}},\nb\\
&&A_{s+1}(\xi)=-\frac{1}{2} B'_s(\xi)+\frac{1}{2} \int \psi(v) B_s(v) dv,\nb\\
\eqn
where $\pm$ correspond to $\xi \geq 0$ and $\xi\leq 0$, respectively.

Similarly, for the branches $\text{Bi}(\lambda^{2/3}\xi)$, we  find
\bqn
U(\lambda,\xi)=\hat A(\lambda, \xi)\text{Bi}(\lambda^{2/3}\xi)+\lambda^{-4/3}\hat B(\lambda, \xi) \text{Bi}'(\lambda^{2/3}\xi).\nb\\
\eqn
It is easy to check that $\hat A(\lambda, \xi)$ and $\hat B(\lambda, \xi)$ have the same expressions as those  given in  Eq.(\ref{AB}). Then, the approximate solution of $U(\xi)$ up to the $(2n)$-th order of the approximation can be expressed as \cite{Olver1974,uniform_singleH}
\bqn\lb{Uy0}
U(\lambda,\xi)&=&\alpha_0 \Bigg[\text{Ai}(\lambda^{2/3} \xi) \sum_{s=0}^{n} \frac{A_s(\xi)}{\lambda^{2s}}\nb\\
&&~~~~~~~~+\frac{\text{Ai}'(\lambda^{2/3}\xi)}{\lambda^{4/3}} \sum_{s=0}^{n-1} \frac{B_s(\xi)}{\lambda^{2s}}+\epsilon_3^{(2n+1)}\Bigg]\nb\\
&&+\beta_0 \Bigg[\text{Bi}(\lambda^{2/3} \xi) \sum_{s=0}^{n} \frac{A_s(\xi)}{\lambda^{2s}}\nb\\
&&~~~~~~~~+\frac{\text{Bi}'(\lambda^{2/3}\xi)}{\lambda^{4/3}} \sum_{s=0}^{n-1} \frac{B_s(\xi)}{\lambda^{2s}}+\epsilon_4^{(2n+1)}\Bigg],\nb\\
\eqn
where the error bounds $\epsilon_3^{(2n+1)}$ and $\epsilon_{4}^{(2n+1)}$ for the $(2n)$-th order  are given by
\bqn
&&\frac{\epsilon_{3}^{(2n+1)}}{M(\lambda^{2/3} \xi)}, \;\;\frac{\partial \epsilon_{3}^{(2n+1)}/\partial\xi}{\lambda^{2/3} N(\lambda^{2/3}\xi)} \nb\\
&&~~~~\leq 2 E^{-1}(\lambda^{2/3}\xi)  \exp{\Big\{\frac{2\kappa_0 \mathscr{V}_{\alpha,\xi}(|\xi^{1/2}|B_0)}{\lambda}\Big\}} \nb\\
&&~~~~~~~~~~~~\times\frac{\mathscr{V}_{\alpha,\xi}(|\xi^{1/2}|B_n)}{\lambda^{2n+1}},\nb\\
&&\frac{\epsilon_{4}^{(2n+1)}}{M(\lambda^{2/3} \xi)}, \;\;\frac{\partial \epsilon_{4}^{(2n+1)}/\partial\xi}{\lambda^{2/3} N(\lambda^{2/3}\xi)} \nb\\
&&~~~~\leq 2 E(\lambda^{2/3}\xi) \exp{\Big\{\frac{2\kappa_0 \mathscr{V}_{\xi,\beta}(|\xi^{1/2}|B_0)}{\lambda}\Big\}} \nb\\
&&~~~~~~~~~~~~\times\frac{\mathscr{V}_{\xi,\beta}(|\xi^{1/2}|B_n)}{\lambda^{2n+1}}.
\eqn

\subsection{\lb{appendixA4}Approximate solution near the turning points $y_1$ and $y_2$}

Except  single turning points, the uniform asymptotic approximation can be also applied to other types of turning points, like high-order turing  or multi-turning points. In this subsection, we shall consider a pair of turning points, $y_1$ and $y_2$, which could be both single and real, coalescent, or complex conjugate. In the neighborhoods of this type of turning points, we have
\bqn
\hat g(y)=p(y)(y-y_1)(y-y_2),
\eqn
where $p(y)$ is regular in the interval of interest, and $p(y_1) \not =0$ and $p(y_2) \not =0$. Without loss of the generality, here we assume that $p(y) <0$ when $y<y_1$ and $y>y_2$ with $y_1<y_2$. In general, we have three different cases depending on the nature of the two turning points $y_1$ and $y_2$, 
\begin{itemize}
\item $y_1$ and $y_2$ are two distinct real roots of $\hat g(y)$, i.e., {\em two single and real turning points} [c.f. Case (a) in Fig. \ref{fig_gofy}];
\item $y_1=y_2$, a double root of $\hat g(y)$, i.e., {\em a double turning point} [c.f. Case (b) in Fig. \ref{fig_gofy}];
\item $y_1$ and $y_2$ are two  complex  roots of $\hat g(y)$, i.e., {\em two complex conjugate turning points} [c.f. Cases (c) and (d) in Fig. \ref{fig_gofy}]. Recall that the difference between Cases (c) and (d)  is that for Case (d), the two complex roots are largely spaced in the imaginary axis.
\end{itemize}
In the uniform asymptotic approximation, we require { the condition } that {\em the function $|q(y)|$ be small compared with $|\lambda^2 \hat g(y)|$, except in the neighborhoods of $ y = y_1$ and $y = y_2$, in which it must be smaller  than $|\lambda^2 \hat g(y) (y-y_1)^{-1} (y-y_2)^{-1}|$} \cite{uniform_nonlinear}. Following Olver  \cite{Olver1975}, we shall adopt a method to treat all of the above  cases together. The crucial point is to choose $f^{(1)}(\zeta)^2$ in the Liouville transformation (\ref{Olver trans}) so that \footnote{Here we use $\zeta$ to denote the variable $\xi(y)$ in the Liouville transformation, for the purpose to distinguish with the variable $\xi(y)$ used for the $y_0$ case.}
\bqn
\lb{xi0A}
f^{(1)}(\zeta)^2=|\zeta^2-\zeta_0^2|,
\eqn
where we choose $\zeta$ to be an increasing function of $y$, and with the conditions $\zeta(y_1)=-\zeta_0$ and $\zeta(y_2)=\zeta_0$. The quantity $\zeta_0^2$ can be positive, zero, and negative, depending on whether  $y_{1,2}$ are both real and different
$y_1 \not= y_2$, or both real but equal $y_1=y_2$, or $y_{1,2}$ are complex conjugate. Then, it can be shown that $\zeta_0^2$ can be expressed as, 
\bqn\lb{zeta_0}
\zeta_0^2=\frac{2}{\pi}\int_{y_1}^{y_2} \sqrt{\hat g(y)}dy.
\eqn

Now let us turn to derive the relation between $\zeta(y)$ and $y$. We first consider the case where $y_1$ and $y_2$ are real and $y>y_2$, so that  $\zeta(y)>\zeta_0$. Then, from Eq.(\ref{Olver trans}) we find
\bqn
\int_{\zeta_0}^{\zeta}\sqrt{v^2-\zeta_0^2}dv=\int_{y_2}^y\sqrt{-\hat g(y')}dy',
\eqn
which yields
\bqn\lb{zeta1}
&&\int_{y_2}^y\sqrt{-\hat g(y')}dy'
=\frac{1}{2}\zeta\sqrt{\zeta^2-\zeta_0^2}-\frac{\zeta_0^2}{2}\operatorname{arcosh}{\left(\frac{\zeta}{\zeta_0}\right)}.\nb\\
\eqn
When $y<y_1$, we have $\zeta(y)<-\zeta_0$, then from Eq.(\ref{Olver trans}) we find
\bqn
\int_{-\zeta_0}^{\zeta}\sqrt{v^2-\zeta_0^2}dv=\int_{y_1}^y\sqrt{-\hat g(y')}dy',
\eqn
which yields
\bqn\lb{zeta2}
&&\int_{y_1}^y\sqrt{-\hat g(y')}dy'
=\frac{1}{2}\zeta\sqrt{\zeta^2-\zeta_0^2}+\frac{\zeta_0^2}{2}\operatorname{arcosh}{\left(-\frac{\zeta}{\zeta_0}\right)}.\nb\\
\eqn
When $y_1\leq y\leq y_2$, we have $-\zeta_0 <\zeta(y)<\zeta_0$, and
\bqn
\int_{-\zeta_0}^\zeta \sqrt{\zeta_0^2-v^2}dv=\int_{y_1}^y \sqrt{\hat g(y')}dy',
\eqn
which yields
\bqn\lb{zeta3}
\int_{y_1}^y \sqrt{\hat g(y')}dy'&=&\frac{1}{2}\zeta\sqrt{\zeta^2-\zeta_0^2}+\frac{\zeta_0^2}{2}\arccos\left(-\frac{\zeta}{\zeta_0}\right).\nb\\
\eqn
Now let us turn to consider the case where  $y_1$ and $y_2$ are complex conjugate. In this case $\zeta_0^2$ is always negative, thus from Eq.(\ref{Olver trans}) one finds
\bqn
\int_0^\zeta \sqrt{\zeta^2-\zeta_0^2}d\zeta=\int_{\text{Re}(y_1)}^y \sqrt{-\hat g(y')}dy',
\eqn
which yields
\bqn\lb{zeta4}
&&\int_{\text{Re}(y_1)}^y \sqrt{-\hat g(y')}dy'\nb\\
&&\;\;\;=\frac{1}{2}\zeta\sqrt{\zeta^2-\zeta_0^2}-\frac{\zeta_0^2}{2}\ln\left(\frac{\zeta+\sqrt{\zeta^2-\zeta_0^2}}{|\zeta_0|}\right).
\eqn
Then,  with $f^{(1)}(\zeta)^2$ given by Eq.(\ref{xi0A}),  Eq.(\ref{eomU}) reduces to
\bqn
\lb{eomy1y2}
\frac{d^2U}{d\zeta^2}=\left[\lambda^2 \left(\zeta_0^2-\zeta^2\right)+\psi(\zeta)\right]U.
\eqn
Neglecting the $\psi(\zeta)$ term, we find that  the approximate solution can be expressed in terms of the parabolic cylinder functions $W(\frac{1}{2}\lambda \zeta_0^2,\pm \sqrt{2\lambda} \zeta)$,
and is given by
\bqn
U(\zeta)&=& \alpha_1 \Bigg\{W\left(\frac{1}{2}\lambda \zeta_0^2, \sqrt{2\lambda}\zeta \right)+\epsilon_5^{(1)}\Bigg\}\nb\\
&&+\beta_1 \Bigg\{W\left(\frac{1}{2}\lambda \zeta_0^2, -\sqrt{2\lambda }\zeta \right)+\epsilon_6^{(1)}\Bigg\},
\eqn
from which we find
\bqn\lb{solutionW}
\mu_k(y)&=&\alpha_1 \left(\frac{\zeta^2-\zeta_0^2}{-\hat g(y)}\right)^{\frac{1}{4}} \left[W\left(\frac{1}{2}\lambda \zeta_0^2, \sqrt{2\lambda}\zeta \right)+\epsilon_5^{(1)}\right]\nb\\
&&+\beta_1 \left(\frac{\zeta^2-\zeta_0^2}{-\hat g(y)}\right)^{\frac{1}{4}} \left[W\left(\frac{1}{2}\lambda \zeta_0^2, -\sqrt{2\lambda }\zeta \right)+\epsilon_6^{(1)}\right],\nb\\
\eqn
where $\epsilon_5^{(1)}$ and $\epsilon_6^{(1)}$ are  the errors of the corresponding first-order approximations, which are bounded by
\bqn
&&\frac{|\epsilon_5^{(1)}|}{M\left(\frac{1}{2}\lambda \zeta_0^2,\sqrt{2\lambda }\zeta\right)},\;\frac{|\partial \epsilon_5^{(1)}/\partial \zeta|}{\sqrt{2} N\left(\frac{1}{2}\lambda \zeta_0^2,\sqrt{2\lambda }\zeta\right)}\nb\\
&&\;\;\;\leq \frac{\kappa}{\lambda_0 E\left(\frac{1}{2}\lambda \zeta_0^2,\sqrt{2\lambda }\zeta\right)} \Bigg\{\exp{\Big(\lambda \mathscr{V}_{\zeta,a_5}(\mathscr{I})\Big)}-1\Bigg\}, ~~~~~~\nb\\
&&\frac{|\epsilon_6^{(1)}|}{M\left(\frac{1}{2}\lambda \zeta_0^2,\sqrt{2\lambda }\zeta\right)},\;\frac{|\partial \epsilon_6^{(1)}/\partial \zeta|}{\sqrt{2} N\left(\frac{1}{2}\lambda \zeta_0^2,\sqrt{2\lambda }\zeta\right)}\nb\\
&&\;\;\;\leq \frac{\kappa E\left(\frac{1}{2}\lambda \zeta_0^2,\sqrt{2\lambda }\zeta\right)}{\lambda } \Bigg\{\exp{\Big(\lambda_0 \mathscr{V}_{a_6,\zeta}(\mathscr{I})\Big)}-1\Bigg\}, ~~~~~~~
\eqn
where $M\left(\frac{1}{2}\lambda \zeta_0^2,\sqrt{2\lambda }\zeta\right)$, $N\left(\frac{1}{2}\lambda \zeta_0^2,\sqrt{2\lambda }\zeta\right)$, and $E\left(\frac{1}{2}\lambda \zeta_0^2,\sqrt{2\lambda }\zeta\right)$ are auxiliary functions of the parabolic cylinder functions which are given in \cite{uniform_nonlinear}, $a_5$ and $a_6$ denotes the upper and lower limit of the variable $\zeta$ respectively, and
\bqn
\mathscr{I}(\zeta) \equiv \int \frac{\psi(\zeta)}{\sqrt{|\zeta^2-\zeta_0^2|}}dv, 
\eqn
is the associated error control function for the approximate solutions near $y_1$ and $y_2$.
 
Now we need to extend the above first-order approximate solution to high orders. Following Olver \cite{Olver1975}, we  assume that the exact solution $U(\lambda,\zeta)$ takes the form (for the branch of  $W(\frac{1}{2}\lambda\zeta_0^2, \sqrt{2\lambda} \zeta)$)
\bqn\lb{Uy1y2}
U(\lambda,\zeta)&=&C(\lambda,\zeta) W\left(\frac{1}{2}\lambda\zeta_0^2, \sqrt{2\lambda} \zeta\right)\nb\\
&&+\frac{\sqrt{2\lambda}}{\lambda^2}D(\lambda,\zeta)W'\left(\frac{1}{2}\lambda\zeta_0^2, \sqrt{2\lambda} \zeta\right),
\eqn
where
\bqn
W'\left(\frac{1}{2}\lambda\zeta_0^2,\sqrt{2\lambda} \zeta\right)\equiv\frac{\partial W(\frac{1}{2}\lambda\zeta_0^2,\sqrt{2\lambda} \zeta)}{\partial(\sqrt{2\lambda}\zeta)}.
\eqn
As a result, we find
\bqn
W''\left(\frac{1}{2}\lambda\zeta_0^2, \sqrt{2\lambda} \zeta\right)=\frac{\lambda}{2} (\zeta_0^2-\zeta^2) W\left(\frac{1}{2}\lambda\zeta_0^2, \sqrt{2\lambda} \zeta\right).\nb\\
\eqn
For the first-order approximation we have $C(\lambda,\zeta)=1$ and $D(\lambda,\zeta)=0$. Substituting Eq.(\ref{Uy1y2}) into the equation of motion and then equating the coefficients of $W(\frac{1}{2}\lambda\zeta_0^2, \sqrt{2\lambda} \zeta)$ and $W'(\frac{1}{2}\lambda\zeta_0^2, \sqrt{2\lambda} \zeta)$, we obtain
\bqn
2 \lambda^2 C'+D''- \psi D=0,\nb\\
C''-2 \zeta D-2  (\zeta^2-\zeta_0^2) D'- \psi C=0.
\eqn
Expanding the functions $C$ and $D$ as 
\bqn
C(\lambda,\zeta)&=&\sum_{s=0}^{+\infty} \lambda^{-2s } C_s(\zeta),\nb\\
D(\lambda,\zeta)&=& \sum_{s=0}^{+\infty} \lambda^{-2s} D_s(\zeta),
\eqn
 we find
\bqn
2  C'_{s+1}+D''_s- \psi D_s=0,\nb\\
C''_s- 2 \zeta D_s-2  (\zeta^2-\zeta_0^2) D'_s- \psi C_s=0.
\eqn
Integration of the above two equations leads to, 
\bqn\lb{AB2}
 C_0(\zeta)&=&1,\;\;\nb\\
D_s(\zeta)&=&\begin{cases}
\frac{1}{2\sqrt{\zeta^2-\zeta_0^2}}\int_{\zeta_0}^\zeta \frac{C''_s(v)-\psi(v) C_s(v)}{\sqrt{v^2-\zeta_0^2}}dv  & |\zeta| \geq |\zeta_0|,\\
\frac{-1}{2\sqrt{\zeta_0^2-\zeta^2}}\int_{\zeta_0}^\zeta \frac{C''_s(v)-\psi(v) C_s(v)}{\sqrt{\zeta_0^2-v^2}}dv  & |\zeta| \leq |\zeta_0|,
\end{cases}\nb\\
C_{s+1}(\zeta)&=&-\frac{1}{2} D'_s(\zeta)+\frac{1}{2} \int^\zeta \psi(v) D_s(v) dv.
\eqn
Similarly, for the branch $W(\frac{1}{2}\lambda\zeta_0^2, -\sqrt{2\lambda} \zeta)$, we  find
\bqn
U(\zeta)&=&\hat C(\lambda,\zeta)W\left(\frac{1}{2}\lambda\zeta_0^2, -\sqrt{2\lambda} \zeta\right)\nb\\
&&-\frac{\sqrt{2\lambda}}{\lambda^{2}}\hat D(\lambda,\zeta) W'\left(\frac{1}{2}\lambda\zeta_0^2, -\sqrt{2\lambda} \zeta\right).
\eqn
Here $\hat C(\lambda,\zeta)$ and $\hat D(\lambda,\zeta)$ have the same expressions as those given in Eq.(\ref{AB2}).
Then,  the approximate solution $U(\zeta)$ up to the $(2n)$-th order can be expressed as
\begin{widetext}
\bqn\lb{Uy1y2}
U(\zeta)&=&\alpha_1 \Bigg[W\left(\frac{1}{2}\lambda\zeta_0^2, \sqrt{2\lambda} \zeta\right) \sum_{s=0}^{n} \frac{C_s(\zeta)}{\lambda^{2s}}+\frac{\sqrt{2\lambda}W'\left(\frac{1}{2}\lambda\zeta_0^2, \sqrt{2\lambda} \zeta\right)}{\lambda^{2}} \sum_{s=0}^{n-1} \frac{D_s(\zeta)}{\lambda^{2s}}+\epsilon_5^{(2n+1)}\Bigg]\nb\\
&&+\beta_1 \Bigg[W\left(\frac{1}{2}\lambda\zeta_0^2, -\sqrt{2\lambda} \zeta\right)\sum_{s=0}^{n} \frac{C_s(\zeta)}{\lambda^{2s}}-\frac{\sqrt{2\lambda}W'\left(\frac{1}{2}\lambda\zeta_0^2, -\sqrt{2\lambda} \zeta\right)}{\lambda^{2}} \sum_{s=0}^{n-1} \frac{D_s(\zeta)}{\lambda^{2s}}+\epsilon_6^{(2n+1)}\Bigg],
\eqn
with
\bqn
&&\frac{|\epsilon_5^{(2n+1)}|}{M(\frac{1}{2}\lambda\zeta_0^2,\sqrt{2\lambda}\zeta)},\; %
\frac{|\partial \epsilon_5^{(2n+1)}/\partial \zeta|}{\sqrt{2\lambda} N(\frac{1}{2}\lambda \zeta_0^2,\sqrt{2\lambda}\zeta)}\leq \frac{\kappa \exp{\Big(\frac{\kappa_0}{\lambda} \mathscr{V}_{\zeta,a_5}(\sqrt{|\zeta^2-\zeta_0^2|}D_0)\Big)}}{ E(\frac{1}{2}\lambda\zeta_0^2,\sqrt{2\lambda }\zeta)}
\frac{\mathscr{V}_{\zeta,a_5}(\sqrt{|\zeta^2-\zeta_0^2|}D_n)}{\lambda^{2n+1}},\nb\\
&&\frac{|\epsilon_6^{(2n+1)}|}{M(\frac{1}{2}\lambda\zeta_0^2,\sqrt{2\lambda}\zeta)},\; %
\frac{|\partial \epsilon_6^{(2n+1)}/\partial \zeta|}{\sqrt{2\lambda} N(\frac{1}{2}\lambda \zeta_0^2,\sqrt{2\lambda}\zeta)}\leq \frac{\kappa \exp{\Big(\frac{\kappa_0}{\lambda} \mathscr{V}_{a_6,\zeta}(\sqrt{|\zeta^2-\zeta_0^2|}D_0))\Big)}}{ E^{-1}(\frac{1}{2}\lambda\zeta_0^2,\sqrt{2\lambda }\zeta)}
\frac{\mathscr{V}_{a_6,\zeta}(\sqrt{|\zeta^2-\zeta_0^2|}D_n)}{\lambda^{2n+1}}.
\eqn
where $\epsilon_5^{(2n+1)}$ and $\epsilon_6^{(2n+1)}$ are errors of the $(2n)$-th order approximation. 
\end{widetext}

\section{\lb{appendixB}Matching of the approximate solutions}
\renewcommand{\theequation}{B.\arabic{equation}} \setcounter{equation}{0}

%
\subsection{Matching of the approximate solutions}

Now we need to match   the individual solutions obtained above together. The first step is to match the approximate solution associated with turning points $y_1$ and $y_2$ with the following initial condition,
\bqn\lb{ad_vacuum}
\lim_{y\rightarrow+\infty}\mu_k(y)&=&\frac{1}{\sqrt{2 \omega}} e^{-i\int \omega d\eta}\nb\\
&\simeq& \sqrt{\frac{1}{2\lambda k}}\frac{1}{(-\hat{g})^{1/4}} \exp{\left(i \lambda\int_{y_i}^y \sqrt{-\hat{g}}dy\right)}.\nb\\
\eqn
However, the approximate solution involves so many high-order terms, which make the matching very complicated. In order to simplify it, let us study their behavior in the limit $y\rightarrow +\infty$. Let us first consider the $D_0(\zeta)$ term in  Eq.(\ref{Uy1y2}), which is given by
\bqn
D_0(\zeta)&=&-\frac{1}{2\sqrt{\zeta^2-\zeta_0^2}} \int_{\zeta_0}^{\zeta} \frac{\psi(v)}{\sqrt{v^2-\zeta_0^2}}dv\nb\\
&=&-\frac{\mathscr{I}(\zeta)}{2\sqrt{\zeta^2-\zeta_0^2}}.
\eqn
Note that in the above expression we had used $C_0(\zeta)=1$, where $\mathscr{I}(\zeta)$ is the error control function associated with the approximate solution around $y_1$ and $y_2$, which behaves well around these two turning points. The integrand in the error control function can be expressed as
\bqn
\frac{\psi(v)}{\sqrt{v^2-\zeta_0^2}} &=&\Bigg[\frac{q}{\hat{g}}-\frac{5 (\hat{g}')^2-4 \hat{g} \hat{g}''}{16\hat{g}^3}\nb\\
&&\;\;+\frac{5\zeta_0^2}{4(v^2-\zeta_0^2)^3}+\frac{3}{4(v^2-\zeta_0^2)^2}\Bigg] \sqrt{v^2-\zeta_0^2},\nb\\
\eqn
and using $\sqrt{v^2-\zeta_0^2}dv=\sqrt{-\hat{g}}dy$,  we find that
\bqn
\mathscr{I}(\zeta)=\mathscr{F}(\zeta)+\int_{\zeta_0}^{\zeta} \left[\frac{5\zeta_0^2}{4 (v^2-\zeta_0^2)^{5/2}}+\frac{3}{4(v^2-\zeta_0^2)^{3/2}}\right]dv,\nb\\
\eqn
where $\mathscr{F}(\zeta)$ is the associated error control function of the Liouville-Green (LG) approximate solution near the pole $y = +\infty$ \cite{uniform_nonlinear}. Obviously in the limit $y\rightarrow +\infty$, we have $\mathscr{I}(\zeta)\rightarrow \mathscr{F}(\zeta)$, which has been already proved to be   convergent. As a result, we have
\bqn
\lim_{y\rightarrow +\infty} D_0(\zeta) =- \frac{\mathscr{I}(+\infty)}{2\sqrt{\zeta^2-\zeta_0^2}}.
\eqn
Then, let us  turn to the next order, the term $C_1$, which is given by
\bqn
C_1(\zeta)=-\frac{1}{2} D_0'(\zeta)+\frac{1}{2}\int_{\zeta_0}^{\zeta} \psi(v)D_0(v)dv.
\eqn
In the limit $y\rightarrow +\infty$,  $D_0'(\zeta)$ becomes negligible, and we find
\bqn
\lim_{y\rightarrow+\infty} C_1(\zeta) &=&-\frac{1}{2} \int_{\zeta_0}^{\zeta} \frac{\psi(v)}{\sqrt{v^2-\zeta_0^2}} \left[\frac{1}{2}\int_{\zeta_0}^{v} \frac{\psi(u)}{\sqrt{u^2-\zeta_0^2}} du\right] dv\nb\\
&=&-\frac{1}{2} \left[\frac{\mathscr{I}(+\infty)}{2}\right]^2.
\eqn
In writing down the above expression we have used the formula
\bqn
&&n! \int_{\zeta_0}^{\zeta} f(\zeta_n) \int_{\zeta_0}^{\zeta_n} f(\zeta_{n-1})\cdots \int_{\zeta_0}^{\zeta_2}f(\zeta_1)d\zeta_1d\zeta_2\cdots d\zeta_n\nb\\
&&\;\;\;~~~~~~~~~~~~=\left[\int_{\zeta_0}^{\zeta} f(v)dv\right]^n.
\eqn
Thus,  up to the third-order, we have
\bqn\lb{asy-ABy2}
C(\lambda,\zeta)&=&1-\frac{1}{2\lambda^2} \left[\frac{\mathscr{I}(+\infty)}{2}\right]^2+\mathcal{O}\left(\frac{1}{\lambda^3}\right),\;\;\;\;\;\nb\\
\frac{D(\lambda,\zeta)}{\lambda}&=& -\frac{1}{\sqrt{\zeta^2-\zeta_0^2}} \frac{\mathscr{I}(+\infty)}{2\lambda}+\mathcal{O}\left(\frac{1}{\lambda^3}\right).
\eqn

Using the asymptotic forms of the parabolic cylinder functions presented in Appendix A, on the other hand, we find, 
\bqn\lb{asy-sol}
&&\lim_{y\rightarrow +\infty}\mu_k(y)\nb\\
&&\;\;\;\;=\left(\frac{-1}{\lambda \hat{g}}\right)^{\frac{1}{4}} \Bigg\{(2j^{2})^{1/4} \alpha_1 \left(C \cos{\mathfrak{D}}-\frac{\bar{D}}{\lambda} \sin{\mathfrak{D}}\right)\nb\\
&&\;\;\;\;\;\;\;\;\;\;\;\;\;\;\;\;\;\;\;\;\;\;\;+(2j^{-2})^{1/4} \beta_1 \left(C \sin{\mathfrak{D}}+\frac{\bar{D}}{\lambda} \cos{\mathfrak{D}}\right)\Bigg\},\nb\\
\eqn
where $\bar D \equiv \sqrt{\zeta^2-\zeta_0^2} D$, $j=j(\sqrt{\lambda}\zeta_0)$, and
\bqn
\mathfrak{D} &\equiv& \frac{\lambda}{2}\zeta\sqrt{\zeta^2-\zeta_0^2}-\frac{\lambda}{2}\zeta_0^2\ln\left(\frac{\zeta+\sqrt{\zeta^2-\zeta_0^2}}{\zeta_0}\right)\nb\\
&&+\frac{\pi}{4}+\phi\left(\frac{\lambda}{2}\zeta_0^2\right)\nb\\
&=&\lambda \int_{y_2}^y \sqrt{-\hat g(y')}dy'+\frac{\pi}{4}+\phi\left(\frac{\lambda}{2}\zeta_0^2\right).
\eqn
Here the function $\phi(x)$ is given by Eq.(\ref{phi}). Note that in order to get  Eq.(\ref{asy-sol}), we have ignored all the high order terms in the asymptotic expansions of the parabolic cylinder functions, as they  all become negligible  in the limit $y\rightarrow +\infty$. Now comparing the above solution  with the initial condition, we obtain
\bqn
\alpha_1&=& \frac{\lambda^{-1/4}}{[2j^2(\sqrt{\lambda}\zeta_0)]^{1/4}} \frac{1}{\sqrt{2k}} \frac{e^{-i X}}{C+i \bar{D}/\lambda},\nb\\
\beta_1&=&  \frac{\lambda^{-1/4}}{[2 j^{-2}(\sqrt{\lambda}\zeta_0)]^{1/4}} \frac{i}{\sqrt{2k}} \frac{e^{-i X}}{C+i \bar{D}/\lambda},\nb\\
\eqn
where the irrelevant phase factor $X$ reads
\bqn
X\equiv \lambda \int_{y_i}^{y_2} \sqrt{-\hat g(y')}dy'+\frac{\pi}{4}+\phi\left(\frac{\lambda}{2}\zeta_0^2\right).
\eqn
From  Eq.(\ref{asy-ABy2}) we conclude
\bqn
C+i \frac{\bar{D}}{\lambda} &=& \sqrt{C^2+\bar{D}^2/\lambda^2}e^{i\theta} \nb\\
&=&(1+\mathcal{O}(1/\lambda^3))e^{i\theta},
\eqn
thus the coefficients $\alpha_1$ and $\beta_1$ are given by, 
\bqn
\alpha_1&\simeq&\frac{\lambda^{-1/4}}{[2 j^2(\sqrt{\lambda}\zeta_0)]^{1/4}} \frac{e^{-i (X+\theta)}} {\sqrt{2k}},\nb\\
\beta_1&\simeq& \frac{i\lambda^{-1/4}}{[2j^{-2}(\sqrt{\lambda}\zeta_0)]^{1/4}} \frac{e^{-i (X+\theta)}}{\sqrt{2k}}.
\eqn

Now we turn to match  the approximate solution around $y_1$ and $y_2$ with the one around $y_0$. For the approximate solution around $y_1$ and $y_2$, when $\zeta\ll-|\zeta_0|$ (i.e., $y\ll y_1$), one has
\bqn\lb{asy2-expansion}
\mu_k(y)&\simeq&\left( \frac{-1}{\lambda \hat g}\right)^{1/4}\Bigg\{\alpha_1(2j^{-2})^{1/4}\left( C \sin{\mathfrak{D}}-\frac{\bar D}{\lambda}\cos{\mathfrak{D}} \right)\nb\\
&&~~~~~~~~~\;\;\;\;+\beta_1 (2j^2)^{1/4} \left(C \cos{\mathfrak{D}}+\frac{\bar D}{\lambda} \sin{\mathfrak{D}}\right)\Bigg\}.\nb\\
\eqn
Here
\bqn
\mathfrak{D} &\equiv& -\frac{\lambda}{2}\zeta \sqrt{\zeta^2-\zeta_0^2}-\frac{\lambda}{2}\zeta_0^2 \ln{\left(\frac{\sqrt{\zeta^2-\zeta_0^2}-\zeta}{\zeta_0}\right)}\nb\\
&&\;\;+\frac{\pi}{4}+\phi\left(\frac{\lambda}{2}\zeta_0^2\right)\nb\\
&=&-\lambda \int_{y_1}^y \sqrt{-\hat g}dy'+\frac{\pi}{4}+\phi\left(\frac{\lambda}{2}\zeta_0^2\right).
\eqn
Similar to the case when $y\to +\infty$ ($\zeta \gg |\xi_0|$), we  assume that the turning points $y_0$ and $y_1$ are large spaced.
Thus the coefficients  $C=C(\lambda,\zeta)$ and $D=D(\lambda,\zeta)$ are given by
\bqn
C(\lambda,\zeta) &\simeq& 1-\frac{\mathscr{I}^2(\zeta)}{8\lambda^2}+\mathcal{O}\left(\lambda^{-3}\right),\nb\\
\frac{D(\lambda,\zeta)}{\lambda}&\simeq& -\frac{\mathscr{I}(\zeta)}{2\lambda}+\mathcal{O}\left(\lambda^{-3}\right),
\eqn
with $\mathscr{I}(\zeta)$ given by
\bqn
\mathscr{I}(\zeta)=\int_{-\zeta_0}^{\zeta}\frac{\psi(v)}{\sqrt{v^2-\zeta_0^2}}.
\eqn

Then let us turn to consider the approximate solution near $y_0$. When $\xi$ is large enough (i.e., $y\gg y_0$), using the asymptotic expansions of Airy type functions given  in Appendix A, we have
\bqn\lb{asy1-expansion}
\mu_k(y)&=&\left(\frac{-1}{\lambda^{2/3}\hat g}\right)^{1/4} \Bigg\{\frac{\alpha_0}{\sqrt{\pi}}\left(A \cos{\mathfrak{A}}+\frac{{\bar  B}}{\lambda}\sin{\mathfrak{A}}\right)\nb\\
&&~~~~~~~~~~+\frac{\beta_0}{\sqrt{\pi}}\left(-A \sin{\mathfrak{A}}+\frac{\bar B}{\lambda}\cos{\mathfrak{A}}\right)\Bigg\},\nb\\
\eqn
where $\bar B\equiv \sqrt{-\xi} B(\lambda,\xi)$ and
\bqn
\mathfrak{A} &\equiv& \frac{2}{3}\lambda (-\xi)^{3/2}-\frac{\pi}{4}=\lambda \int_{y_0}^{y} \sqrt{-\hat g} dy'-\frac{\pi}{4}.
\eqn
Up to the third-order approximation, one finds \cite{uniform_singleH}
\bqn\lb{ABcc}
A(\lambda,\zeta) &\simeq& 1-\frac{\mathscr{H}^2(\xi)}{8\lambda^2}+\mathcal{O}(\lambda^{-3}),\nb\\
\frac{\bar B(\lambda,\zeta)}{\lambda}&\simeq & -\frac{\mathscr{H}(\xi)}{2\lambda}+\mathcal{O}\left(\lambda^{-3}\right),
\eqn
with
\bqn
\mathscr{H}(\xi)=\int_{0}^{\zeta} \frac{\psi(v)}{\sqrt{|v|}}dv.
\eqn
Now combining Eqs.(\ref{asy2-expansion}) and (\ref{asy1-expansion}) we obtain
\begin{widetext}
\bqn\lb{alpha_beta_0}
\alpha_0&\simeq& \frac{\sqrt{\pi}}{\lambda^{1/12}} \Bigg\{\big[(2j^2)^{1/4} \beta_1 \cos{\mathfrak{B}}+(2j^{-2})^{1/4}\alpha_1 \sin{\mathfrak{B}}\big]+\frac{\mathscr{I}(\zeta)+\mathscr{H}(\xi)}{2\lambda}\big[(2j^{-2})^{1/4} \alpha_1 \cos{\mathfrak{B}}-(2j^{2})^{1/4}\beta_1 \sin{\mathfrak{B}}\big]\nb\\
&&~~~~~~~~~~~~ -\frac{1}{2}\left(\frac{\mathscr{I}(\zeta)+\mathscr{H}(\xi)}{2\lambda}\right)^2\big[(2j^2)^{1/4} \beta_1 \cos{\mathfrak{B}}+(2j^{-2})^{1/4}\alpha_1 \sin{\mathfrak{B}}\big]\Bigg\},\nb\\
\beta_0 &\simeq & \frac{\sqrt{\pi}}{\lambda^{1/12}} \Bigg\{\big[(2j^{-2})^{1/4} \alpha_1 \cos{\mathfrak{B}}-(2j^{2})^{1/4}\beta_1 \sin{\mathfrak{B}}\big]-\frac{\mathscr{I}(\zeta)+\mathscr{H}(\xi)}{2\lambda}\big[(2j^{2})^{1/4} \beta_1 \cos{\mathfrak{B}}+(2j^{-2})^{1/4}\alpha_1 \sin{\mathfrak{B}}\big]\nb\\
&&~~~~~~~~~~~~-\frac{1}{2}\left(\frac{\mathscr{I}(\zeta)+\mathscr{H}(\xi)}{2\lambda}\right)^2\big[(2j^{-2})^{1/4} \alpha_1 \cos{\mathfrak{B}}-(2j^{2})^{1/4}\beta_1 \sin{\mathfrak{B}}\big]\Bigg\},
\eqn
\end{widetext}
where
\bqn\lb{Bcc}
\mathfrak{B}\equiv \lambda \int_{y_0}^{y_1} \sqrt{-\hat g(y')}dy'+\phi\left(\frac{\lambda}{2}\zeta_0^2\right),
\eqn
and
\bqn
\mathscr{I}(\zeta)+\mathscr{H}(\xi)=\int_{-\zeta_0}^{\zeta(y)}\frac{\psi(v)}{\sqrt{v^2-\zeta_0^2}}dv+\int_{0}^{\xi(y)}\frac{\psi(v)}{\sqrt{-v}}dv.\nb\\
\eqn
It should be noted that, in order to match the high-order approximate solutions, one has to choose a point at which the two approximate solutions given, respectively, by Eq.(\ref{asy2-expansion}) and Eq.(\ref{asy1-expansion}), are matched. This is unlike the case in the first-order approximation, for which the solutions can be matched at any point between the turning points $y_0$ and $y_1$ (or Re$(y_1)$ when $y_1$ is complex), provided that $\xi(y)$ and $\zeta^2(y)-\zeta_0^2$ both are large enough. While different matching points may lead to  different results,  one can employ the following way to reduce the errors. When we match the solution, we have used both the asymptotic expansions of Airy functions and parabolic cylinder functions, for example
\bqn
\text{Ai}(\lambda^{2/3}\xi)\simeq \frac{\lambda^{-1/6}}{\pi^{1/2}(-\xi)^{1/4}} \left[\cos{\mathfrak{A}}+\mathcal{O}(\xi^{-3/2})\right],\nb\\
\eqn
for $|\xi|\gg 1$ and
\bqn
&&W(\text{\textonehalf}\lambda\zeta_0^2,\sqrt{2\lambda}\zeta) \nb\\
&&\;\;\;\;\simeq \left(\frac{ 2 j^2(\sqrt{\lambda}\zeta_0)}{\lambda (\zeta^2-\zeta_0^2)}\right)^{1/4} \left[\cos{\mathfrak D}+\mathcal{O}\left(\frac{1}{\zeta^2-\zeta_0^2}\right)\right]\nb\\
\eqn
for $\zeta^2-\xi_0^2 \gg 1$. Thus, from the terms we have ignored in the above expansions, it seems that a good matching point should be around the range when
\bqn
\frac{1}{|\xi|^{3/2}} \sim \frac{1}{\zeta^2-\zeta_0^2}.
\eqn
In this paper, we shall always consider the matching when the above condition is satisfied.

\subsection{Error analysis and comparison with numerical (exact) solution}

In this subsection, we are going to study  the above matched approximate solutions   numerically in order to understand the matching process. We also provide the comparison between the approximate solutions and the numerical (exact) ones for several cases.

Let us first focus on the error control functions of the approximate solutions in different ranges. The error control functions  $\mathscr{I}(\zeta)$ and $\mathscr{H}(\xi)$ are presented in Fig. \ref{Fig1}. The left panel corresponds to the error control function  $\mathscr{H}(\xi)$  around the turning point $y_0$, and the right panel corresponds to the error control function $\mathscr{I}(\zeta)$ around turning points $y_1$ and $y_2$. Although the error control function is not the exact error of the approximate solution, it can help us to understand the level of the errors qualitatively. Fig. \ref{Fig1} clearly shows that the error of the approximate solution peaks in the region where $y\to 0^+$. And in a de-Sitter background with $\nu=3/2$, as we have shown in \cite{uniform_singleH}, the error control function $\mathscr{H}(\xi)$ goes to $1/9$ in the limit $y\to 0^+$. Another feature is that $\mathscr{H}(\xi)$ is not sensitive to the values of parameters $b_1$ and $b_2$, as well as $\epsilon_*$. For the error control function $\mathscr{I}(\zeta)$,  unlike $\mathscr{H}(\xi)$, Fig. \ref{Fig1} shows that $\mathscr{I}(\zeta)$ is sensitive to the value of $\epsilon_*$, and for most cases it is extremely small compared to the value of $\mathscr{H}(\xi)$ as $y\to 0^+$. It is clear that $\mathscr{I}(\zeta)$ increases when the parameter $\epsilon_*$ increases.

\begin{figure*}
{
\includegraphics[width=8.1cm]{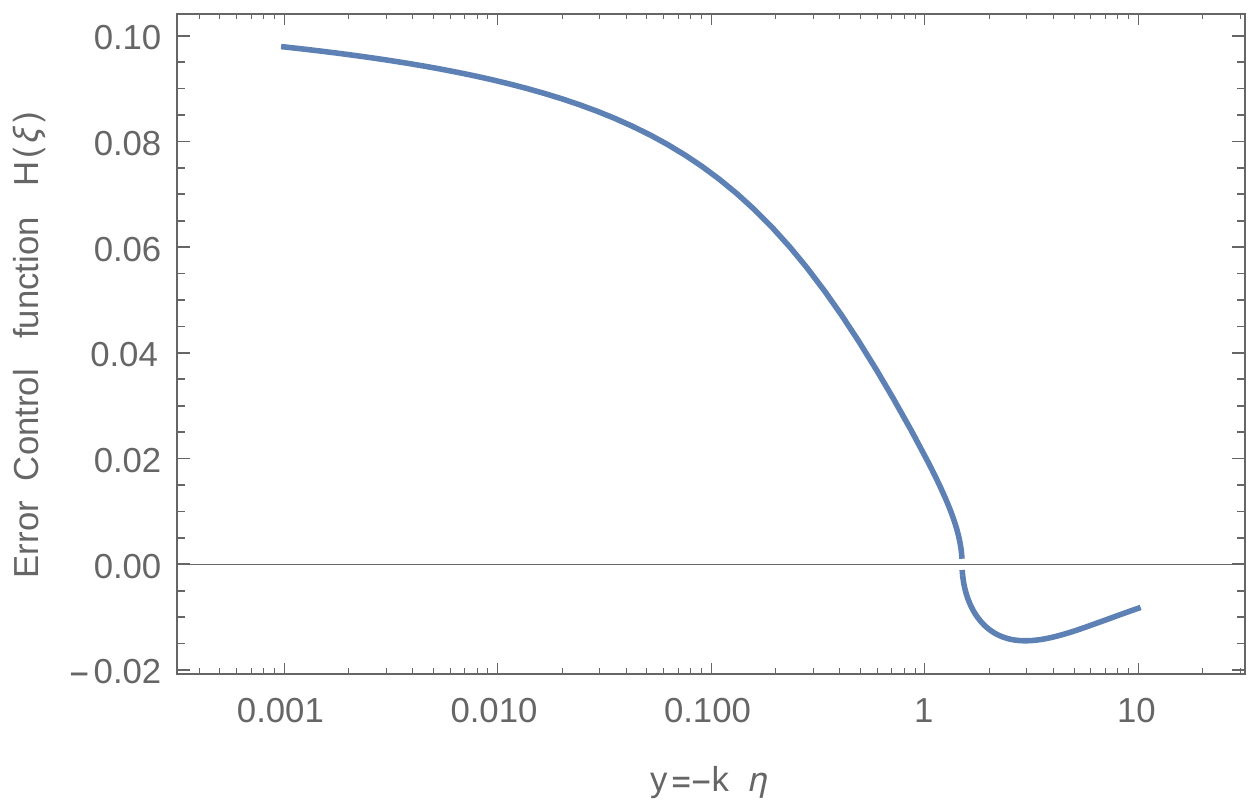}}
{
\includegraphics[width=8.1cm]{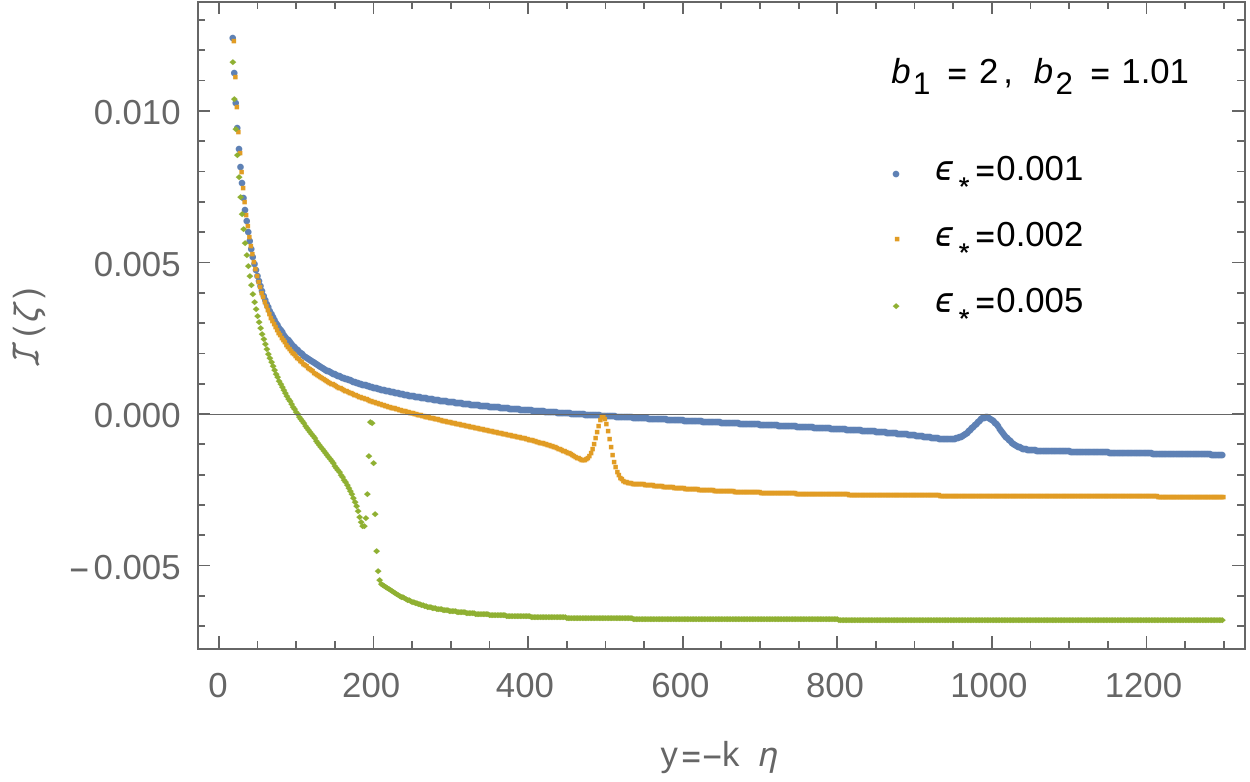}}
\caption{Time evolution of the error control function. (a) Left panel: the error control function $\mathscr{H}(\xi)$ as a function of $y$ in a de-Sitter background. (b) Right panel: the error control function $\mathscr{I}(\zeta)$ as a function of $y$ in the de-Sitter background.}\lb{Fig1}
\end{figure*}

The coefficients $A_1(\xi)$ and $B_0(\xi)$ associated with the Airy  solutions are displayed in Fig. \ref{Fig2}. The left panel corresponds to the evolution of $A_1(\xi)$ as a function of $y$, and the right panel corresponds to the evolution of $B_0(\xi)$. In this figure,  the de-Sitter background is chosen. For the parabolic cylinder solutions, the corresponding coefficients $C_1(\zeta)$ and $D_0(\zeta)$ are presented in Fig. \ref{Fig3} as a function of $y$, where the left panel corresponds to $C_1(\zeta)$ and the right panel represents $D_0(\zeta)$. Similar to the case for the error control functions given in Fig. \ref{Fig1}, now the coefficients $C_1(\zeta)$ and $D_0(\zeta)$ are extremely small and negligible, compared to the values of $A_1(\xi)$ and $B_0(\xi)$, respectively.

\begin{figure*}
{
\includegraphics[width=8.1cm]{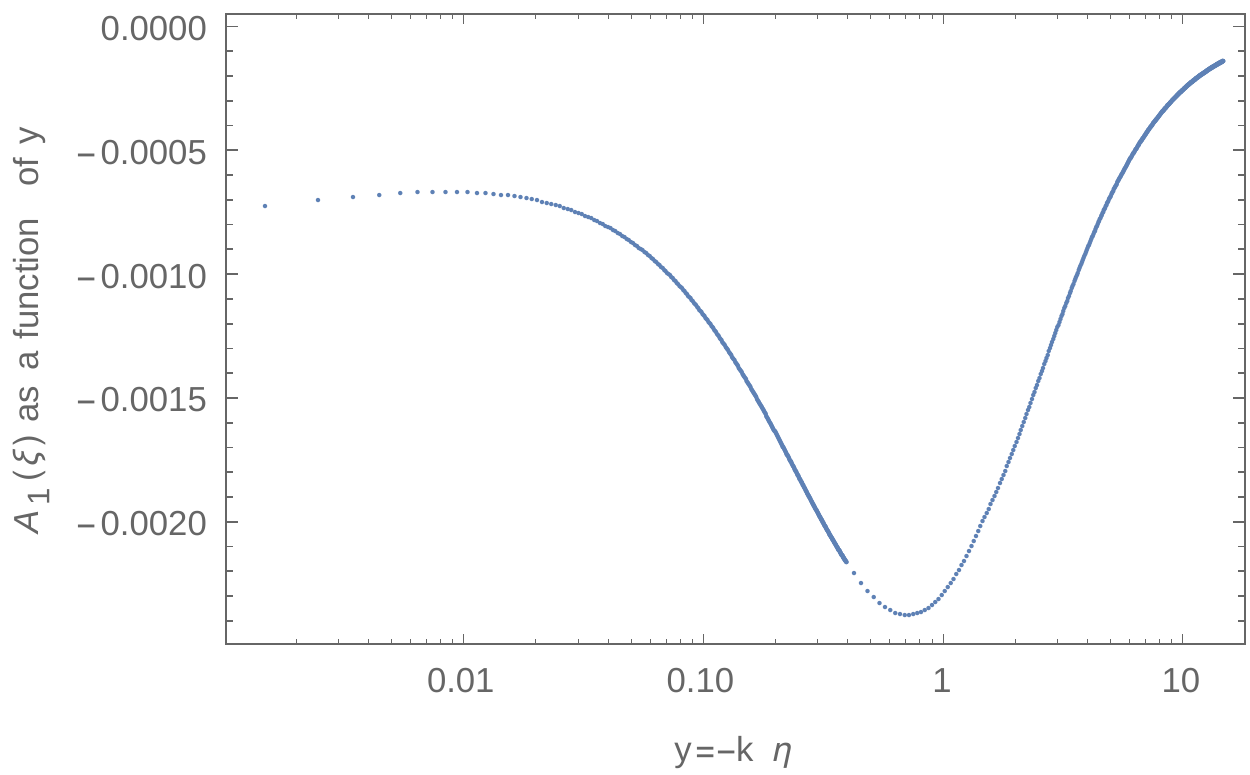}}
{
\includegraphics[width=8.1cm]{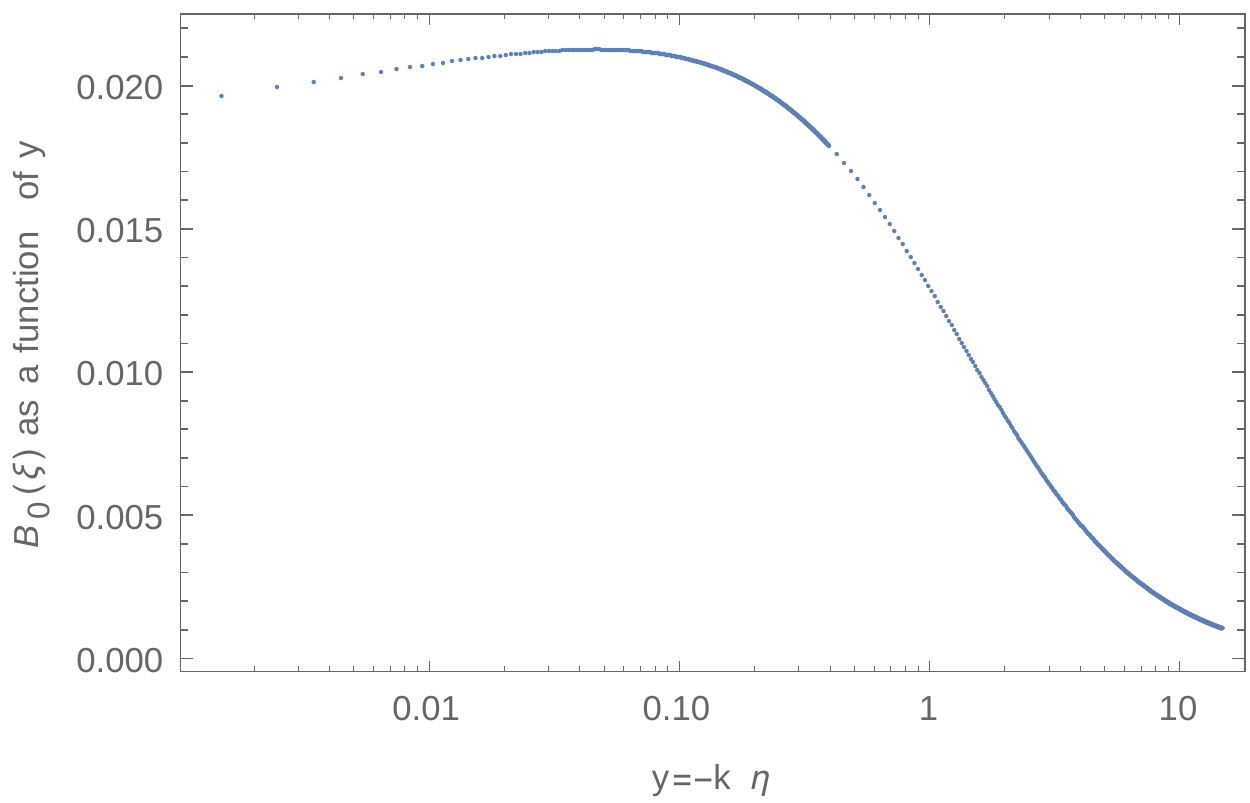}}
\caption{Time evolution of the coefficients $A_1(\xi)$ and $B_0(\xi)$ associated with the Airy type solution around turning points $y_0$. (a) Left panel: $A_1(\xi)$ as a function of $y$ in a de-Sitter background.
(b) Right panel: $B_0(\xi)$ as a function of $y$ in the de-Sitter background.}\lb{Fig2}
\end{figure*}

\begin{figure*}
{\label{C1}
\includegraphics[width=8.1cm]{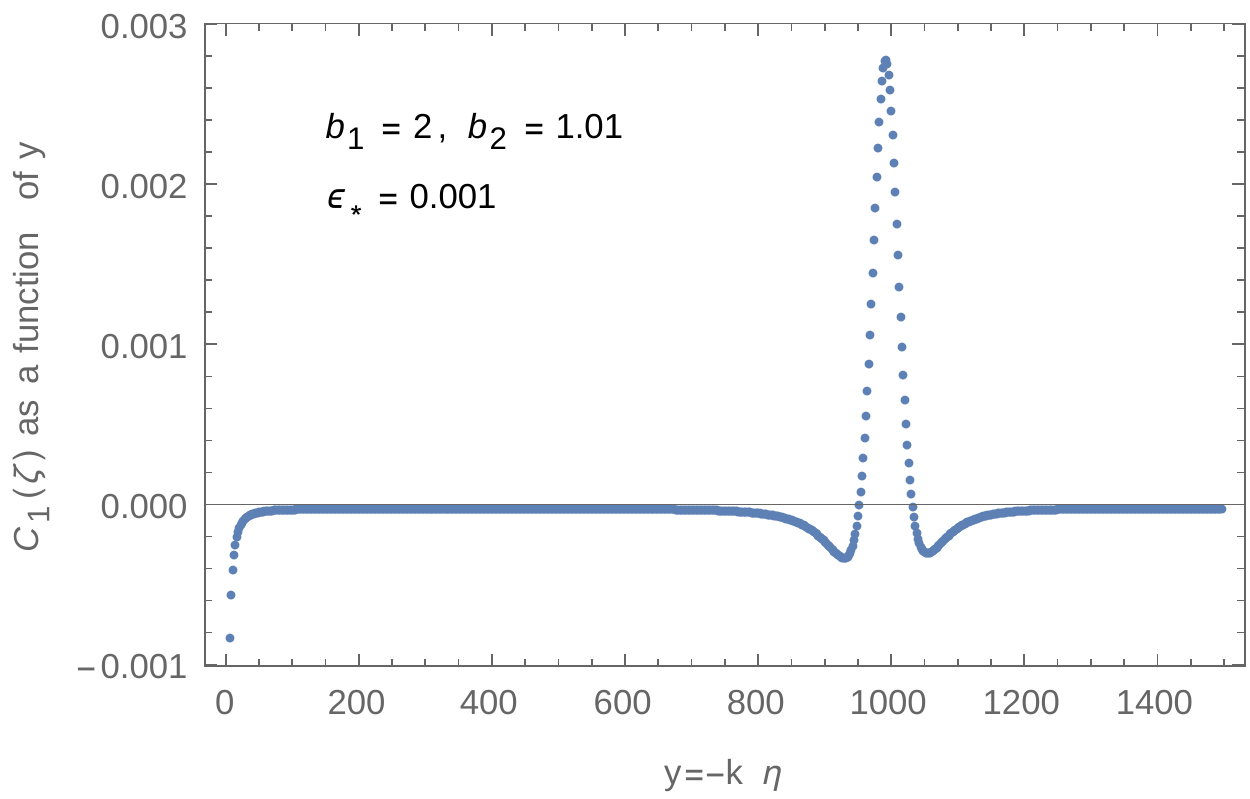}}
{\label{D0}
\includegraphics[width=8.1cm]{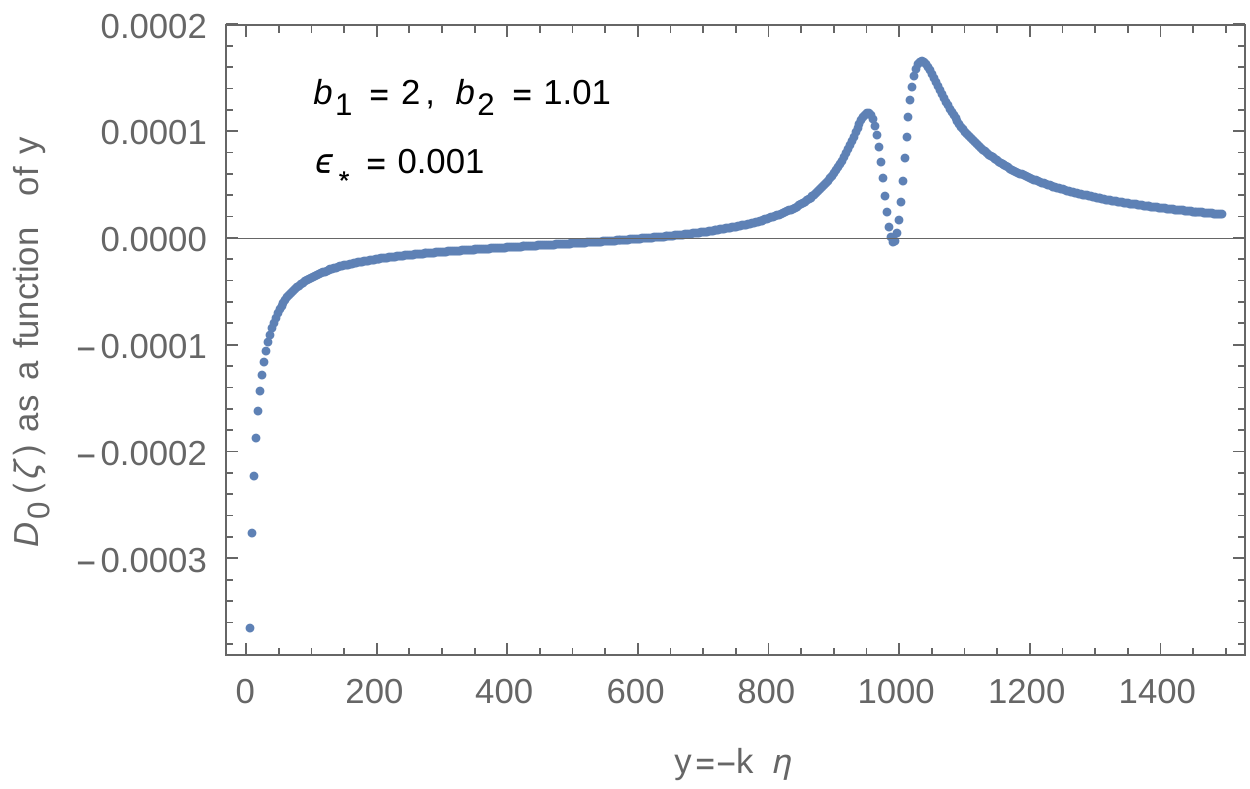}}
\caption{Time evolution of the coefficients $C_1(\zeta)$ and $D_0(\zeta)$ associated with the parabolic cylinder function solutions around turning points $y_1$ and $y_2$. (a) Left panel: $C_1(\zeta)$ as a function of $y$ in a de-Sitter background.
(b) Right panel: $D_0(\zeta)$ as a function of $y$ in the de-Sitter background.} \lb{Fig3}
\end{figure*}

The comparison between the approximate solution and the numerical solution is displayed in Fig. \ref{Fig4}.  As we mentioned  above, the errors of each order of approximation are different when $y\to 0^+$. It has been shown clearly in Fig. \ref{Fig4} that the third-order approximation does improve the precision of the first-order and second-order approximation in the limit $y \to 0^+$. Considering that the error of the third-order approximation is about $0.15\%$, however, the first-order, second-order, and the third-order approximation are almost identical to each other within this error when $y \sim y_1$ or $y \sim y_2$. This implies that, compared to the third-order approximations around the turning point $y_0$, the high-order approximations around the turning points $y_1$ and $y_2$ in general does not significantly improve the first-order approximation.

\begin{figure}
{
\includegraphics[width=8.1cm]{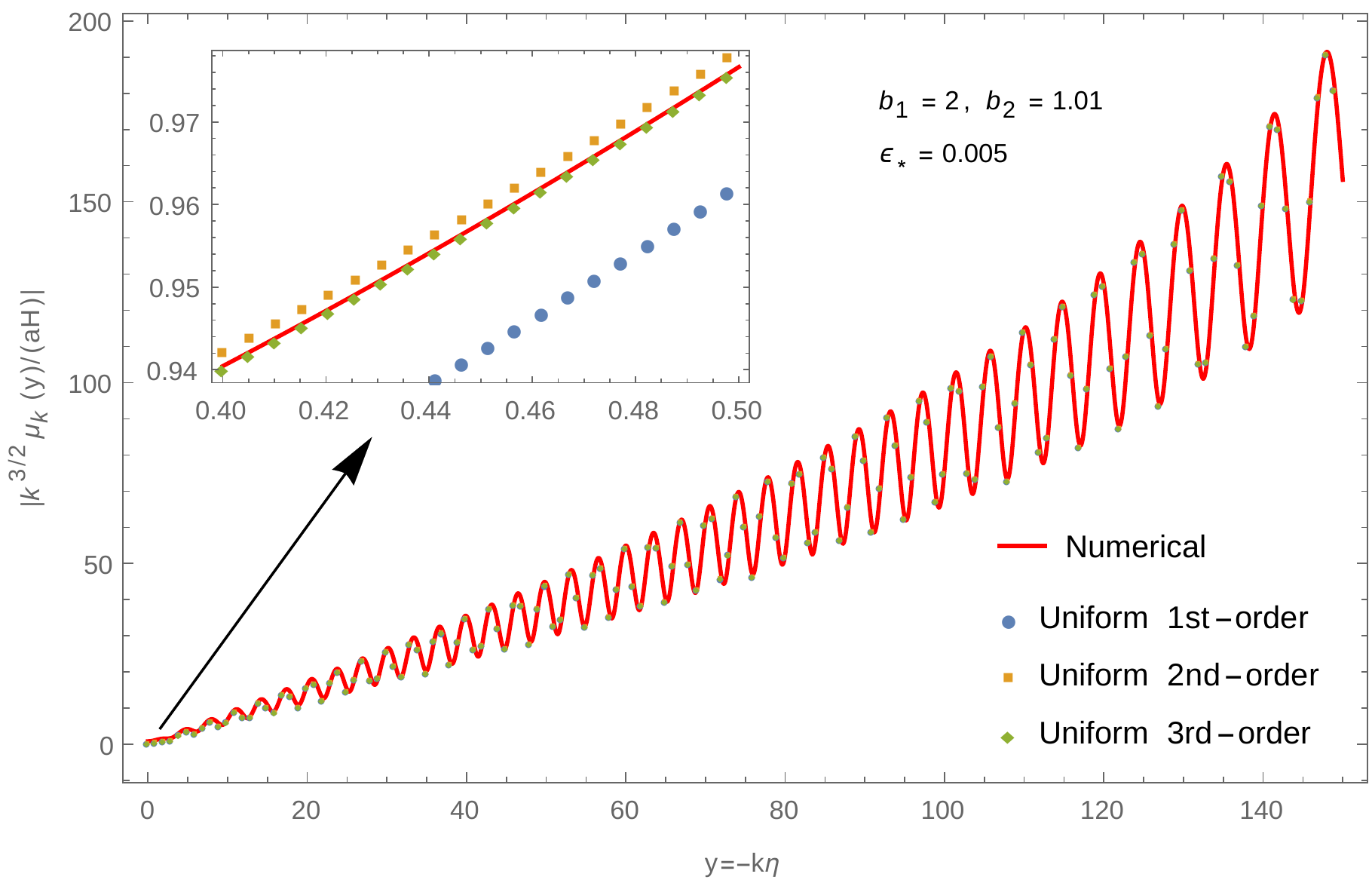}}
\caption{Comparison of  the analytical approximate solutions with different orders of approximations to the numerical (exact) solutions in a de Sitter background.} \lb{Fig4}
\end{figure}

\section{Asymptotical expansions of parabolic cylinder functions and Airy functions}
\renewcommand{\theequation}{C.\arabic{equation}} \setcounter{equation}{0}

The asymptotic expansion of parabolic cylinder functions $W(\text{\textonehalf}\lambda\zeta_0^2,\pm \sqrt{2\lambda}\zeta)$ and $W(\text{\textonehalf}\lambda\zeta_0^2,\pm \sqrt{2\lambda}\zeta)$ for $\zeta^2-\zeta_0^2 \gg 1$ can be written as \cite{Gil2004}
\bqn
W(\text{\textonehalf}\lambda\zeta_0^2,\sqrt{2\lambda}\zeta)&=&\left(\frac{ 2 j^2(\sqrt{\lambda}\zeta_0)}{\lambda (\zeta^2-\zeta_0^2)}\right)^{1/4}\cos{\mathfrak D},\\
W'(\text{\textonehalf}\lambda\zeta_0^2,\sqrt{2\lambda}\zeta)&=&-\left(\frac{\lambda (\zeta^2-\zeta_0^2)}{ 2 j^{-2}(\sqrt{\lambda}\zeta_0)}\right)^{1/4}\sin{\mathfrak D}, ~~~~\\
W(\text{\textonehalf}\lambda\zeta_0^2,-\sqrt{2\lambda}\zeta)&=&\left(\frac{ 2 j^{-2}(\sqrt{\lambda}\zeta_0)}{\lambda (\zeta^2-\zeta_0^2)}\right)^{1/4} \sin{\mathfrak D},\\
W'(\text{\textonehalf}\lambda\zeta_0^2,-\sqrt{2\lambda}\zeta)&=&-\left(\frac{\lambda (\zeta^2-\zeta_0^2)}{ 2 j^{2}(\sqrt{\lambda}\zeta_0)}\right)^{1/4}\cos{\mathfrak D},
\eqn
with
\bqn
\mathfrak D &\equiv& \frac{1}{2} \lambda \zeta \sqrt{\zeta^2-\zeta_0^2} -\frac{1}{2}\lambda \zeta_0^2 \ln{\left(\frac{\zeta+\sqrt{\zeta^2-\zeta_0^2}}{\zeta_0}\right)}\nb\\
&&+\frac{\pi}{4}+\phi\left(\frac{1}{2}\lambda\zeta_0^2\right).
\eqn
Here
\bqn\lb{phi}
\phi(x)&=&\frac{x}{2}-\frac{x}{4}\ln{x^2}+\frac{1}{2}\text{ph}\Gamma\left(\frac{1}{2}+ix\right),
\eqn
where the phase $\text{ph}\Gamma(\frac{1}{2}+i x)$ is zero when $x=0$ and is determined by continuity otherwise.

For Airy type functions $\text{Ai}(x)$ and $\text{Bi}(x)$ for large positive $x$ (i.e., $x \gg 1$), the asymptotic expansions have the form \cite{Olver1974}
\bqn
\text{Ai}(x) &=& \frac{1}{2 \pi ^{1/2} x^{1/4}}e^{- \frac{2}{3} x^{3/2}} ,\nb\\
\text{Bi}(x) &=& \frac{1}{\pi^{1/2} x^{1/4}} e^{\frac{2}{3}x^{3/2}},\nb\\
\text{Ai}(-x) &=& \frac{1}{\pi^{1/2} x^{1/4}} \cos{\left(\frac{2}{3}x^{2/3}-\frac{\pi}{4}\right)}, \nb\\
\text{Bi}(-x) &=& - \frac{1}{\pi^{1/2} x^{1/4}} \sin{\left(\frac{2}{3}x^{2/3}-\frac{\pi}{4}\right)}.
\eqn

\end{document}